\documentclass[journal]{IEEEtran}

\usepackage{amsmath}
\usepackage{amsthm}
\usepackage{epsfig}
\usepackage{graphicx}
\usepackage{epstopdf}
\usepackage{grffile}
\usepackage{times}
\usepackage{url}
\usepackage{color}
\usepackage{cite}
\usepackage{marvosym}
\usepackage{algorithmicx}
\usepackage[ruled]{algorithm}
\usepackage{algpseudocode}
\usepackage{multirow}
\usepackage{stfloats}
\usepackage{amssymb}
\usepackage{algorithmicx}
\usepackage{balance}
\usepackage[ruled]{algorithm}
\usepackage{algpseudocode}

\usepackage[caption=false,font=footnotesize]{subfig}

\newcommand{\new}[1]{{{#1}}}

\begin{document}

\title{Adaptive Mechanism for\\ Distributed Opportunistic Scheduling}

\author{Andres Garcia-Saavedra, Albert Banchs, Pablo Serrano and Joerg Widmer\thanks{A.~Garcia-Saavedra is with Hamilton Institute, Ireland. P.~Serrano is with University Carlos III of Madrid. A.~Banchs is with University Carlos III of Madrid and Institute IMDEA Networks. J.~Widmer is with Institute IMDEA Networks.}
\thanks{This paper is an extended version of our paper \cite{ourinfocom}, which was presented at IEEE INFOCOM 2012.}}
\maketitle

\begin{abstract}
Distributed Opportunistic Scheduling (DOS) techniques have been recently proposed to improve the throughput performance of wireless networks. With DOS, each station contends for the channel with a certain access probability. If a contention is successful, the station measures the channel conditions and transmits in case the channel quality is above a certain threshold. Otherwise, the station does not use the transmission opportunity, allowing all stations to recontend. A key challenge with DOS is to design a distributed algorithm that optimally adjusts the access probability and the threshold of each station. To address this challenge, in this paper we first compute the configuration of these two parameters that jointly optimizes throughput performance in terms of proportional fairness. Then, we propose an adaptive algorithm based on control theory that converges to the desired point of operation. Finally, we conduct a control theoretic analysis of the algorithm to find a setting for its parameters that provides a good tradeoff between stability and speed of convergence. Simulation results validate the design of the proposed mechanism and confirm its advantages over previous proposals.
\end{abstract}

\section{Introduction}
Communication over wireless channels faces two main challenges inherent to the medium: interference and fading. While the former has traditionally been tackled at the MAC layer (for example through techniques such as CSMA/CA and RTS/CTS), the latter has largely been considered \new{as} a physical layer problem (and is usually addressed through proper selection of the transmission rate, i.e., channel coding and modulation scheme). However, the physical layer does not always hide fading effects from the MAC layer \cite{cao2006}, and using very conservative channel coding and modulation schemes that may allow decoding during deep fades wastes capacity.
In contrast, opportunistic scheduling (e.g.,\new{\cite{asadi2013survey, ghosh2009priority}}) addresses the issue of channel quality variations by preferentially scheduling transmissions of senders with good instantaneous channel conditions. Exploiting knowledge of the channel conditions in this manner has been shown to lead to substantial performance gains \new{(e.g., Qualcomm's IS-856)}.
While centralized opportunistic scheduling mechanisms rely on a central entity with global knowledge of the radio conditions of all stations, the more recent Distributed Opportunistic Scheduling (DOS) techniques \cite{itit,ton,twc,infocom, dosjsac}, also work in settings where either such a central entity is not available \new{(e.g., in ad-hoc networks)}, or where the communication overhead to provide timely updates of the channel conditions of all the stations to the central entity is prohibitive \new{(e.g., in case of energy consumption constraints, limited bandwidth, or lack of a control channel)}.  

DOS lets stations contend for channel access and, upon successful contention, a station uses its local information about channel conditions to decide whether to transmit data or give up the transmission opportunity. This decision is taken based on a pure threshold policy, i.e., a station gives up its transmission opportunity if the bit rate allowed by the channel  falls below a certain threshold. By giving up a transmission opportunity and allowing recontention, it is likely that the channel is taken by a station with better channel conditions, resulting in a higher aggregate throughput. Furthermore, since no coordination between stations is required, DOS protocols are simpler to implement and have a lower control overhead compared to centralized approaches.

\new{The seminal work of \cite{itit}} provides valuable insights and a deeper understanding of DOS techniques and their performance. \new{Several works~\cite{ton,twc,infocom, dosjsac} extend the basic mechanism of \cite{itit} to analyze the case of imperfect channel information \cite{twc}, improve channel estimation through two-level channel probing \cite{ton}, and incorporate delay constraints \cite{infocom}. In turn, \cite{dosjsac}  proposes the idea of effective
observation points to avoid the assumption of independent observations during the probing phase used in \cite{itit}. A fundamental drawback of these works is that they \emph{only} aim to maximize  total throughput, an objective that may cause the starvation of those stations with poor link conditions. Heterogeneous links are considered in \cite{capacity-2014} and \cite{dos-hybrid}. The authors of \cite{capacity-2014} study the asymptotic sum-rate capacity of MIMO systems that exploit opportunism with a threshold policy, including non-homogeneous users, which requires some global information (like the number of links contending in the network) and assume a Gaussian channel model; in contrast, our approach  relies on local information only and does not  take any assumption on the distribution of the channel.
The authors of \cite{dos-hybrid}  consider two types of links which may have different QoS constraints but only optimize the thresholds and do not consider non-saturated stations, whereas we jointly optimize access probabilities and thresholds and support different traffic loads. 

The contributions of this paper are the following:
\begin{itemize}
 \item[($i$)] While previous works only optimize the transmission rate thresholds, we perform a joint optimization of both the thresholds and the access probabilities. Our optimization provides a \emph{proportionally fair} allocation \cite{Kelly} that achieves a good tradeoff between total throughput and fairness in heterogeneous topologies. Although the derivation of the optimal configuration follows similar ideas as \cite{doc}, here we use a different approximation which helps us to remove dependencies on global information without compromising performance. 
 \item[($ii$)] The second contribution is the design of ADOS, a light adaptive scheme based on control theory, that drives the system to the optimal point of operation  with the following advantages:
 \begin{itemize}
 \item ADOS performs well in networks with non-saturated stations.\footnote{A saturated station always has data ready for transmission while a non-saturated station may at times have nothing to send.} The analysis and design of previous approaches require the assumption that all stations are always saturated, resulting in overly conservative behavior under non-saturation conditions. In contrast, our approach adapts to the \emph{actual} network load instead of the number of stations, and hence increases the network capacity when there are non-saturated stations.
 \item ADOS adapts the configuration of the system to the dynamics of the environment, such as mobility or stations joining and leaving the network. In contrast, all previous works (including \cite{ourinfocom}) assume static radio conditions and therefore can only be applied in  scenarios with little or no mobility.
 \item ADOS only relies on information that can be observed locally, in contrast to previous approaches which need global information and hence require substantial signaling.
 \end{itemize}
 \item[($iii$)] The third contribution of the paper is the control theoretic analysis of the proposed mechanisms. This analysis guarantees the convergence and stability of the mechanism, and provides a configuration of its parameters that achieves a good tradeoff between stability and speed of convergence. Prior approaches \cite{itit,ton,twc,infocom, dosjsac, capacity-2014,dos-hybrid} do not provide these guarantees.

\end{itemize}


}



This paper extends very substantially the work we recently presented in \cite{ourinfocom}. 
First, we design a new light algorithm to adapt to changing radio conditions. 
\new{Previous approaches, including \cite{ourinfocom}, require to re-compute the threshold with some periodicity which can be computationally very costly (e.g., the iterative algorithm proposed in \cite{itit}, and used in \cite{ourinfocom}, requires solving definite integrals), which precludes a quick adaptation to changes in the channel conditions.}
The proposed adaptive algorithm is based on control theory, like the algorithm designed in \cite{ourinfocom} to adjust the access probability. However, both the design of the algorithm and its analysis are entirely novel, as the conditions that determine the optimal point of operation (and hence the algorithm design to drive the system to this point) as well as the system dynamics (and thus the control theoretic analysis to guarantee an appropriate reaction to changing conditions) are different from~\cite{ourinfocom}. \new{Second, we discuss the implementability of ADOS using off-the-shelf devices in \S\ref{sec-practical}.}
\new{Third}, we significantly extend the performance evaluation of the mechanism:
\begin{itemize}
\item[($i$)] In addition to comparing ADOS to the team-game approach (TDOS) proposed in \cite{itit}, we also compare it against the non-cooperative approach (NDOS) of \cite{itit} and \new{CSMA/CA}, and show that it not only outperforms TDOS, 
but it performs far better than NDOS \new{and CSMA/CA}. This result is very relevant \new{because ADOS, NDOS and CSMA/CA} use only local information \new{whereas} TDOS requires global information (and thus involves substantial signaling).
\item[($ii$)] \new{In addition to analyzing and validating the configuration of the algorithm to adapt the thresholds to changing radio conditions, we also compare its performance  with the algorithm we presented in \cite{ourinfocom} for a  mobile scenario with different speeds and number of stations.}
\item[($iii$)] We evaluate the proposed algorithm under different load conditions and show that the gains obtained with the proposed approach are even higher than those given in \cite{ourinfocom} when the load of non-saturated stations is small.
\item[($iv$)] \new{We assess the performance of all the mechanisms in the presence of channel estimation errors and show that ADOS outperforms all other approaches in this case too.}
\end{itemize}


The rest of the paper is organized as follows. \S\ref{sec-configuration} presents the analysis of our DOS system and optimizes its configuration in terms of proportional fairness. \S\ref{sec-adaptive} proposes a novel adaptive mechanism, \emph{Adaptive Distributed Opportunistic Scheduling} (ADOS), that drives the system to the configuration obtained previously. ADOS is analyzed in \S\ref{sec-analysis} from a control theoretic standpoint to derive a configuration of the mechanism that provides a good tradeoff between stability and reaction to changes. Its performance is validated via simulations in \S\ref{sec-performance}. \new{\S\ref{sec-practical} explains how to implement ADOS with commodity devices.} Finally, \S\ref{sec-conclusions} concludes the paper.

\section{DOS Optimal Configuration}\label{sec-configuration}

In the following, we compute the optimal configuration of the access probabilities and transmission rate thresholds of a DOS system for a proportionally fair throughput allocation, \new{which is a well known allocation criterion to provide a good tradeoff between maximizing total throughput (which may be unfairly distributed among stations) and a purely fair allocation (that may waste capacity)~\cite{Kelly}. }
While the analysis conducted in this section assumes saturation conditions, the mechanism that we devise in the next section also takes into account the non-saturated case.

\subsection{System Model}

\begin{figure}[t!]
\vspace{-5mm}
\centering
\includegraphics[width=\linewidth]{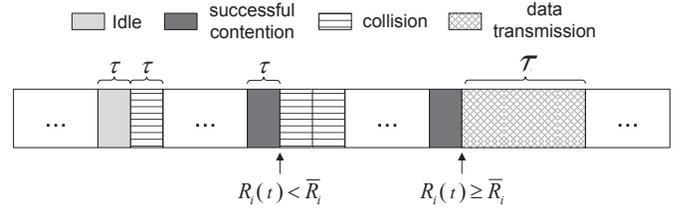}

\caption{\new{An example of the operation of the DOS protocol. The first transmission opportunity  is skipped due to a low available $R_i(t)$ while the second opportunity is used to transmit data due to good channel conditions.}}

\label{fig:system-model}
\end{figure}

Similarly to \cite{itit,ton,twc,infocom, doc}, we model our system as a single-hop contention-based wireless network with $N$ stations where time is divided into mini slots of fixed duration $\tau$. At the beginning of each slot, station $i$ contends for channel access with a given channel access probability, $p_i$. A slot can be empty if none of the stations attempt to access the channel. If $N>1$ stations access the channel in the same slot, a collision occurs and the channel is freed for the next slot. There is a successful contention  if only one station accesses the medium, which then probes the channel. After this channel probing (which we assume takes one slot), the station has perfect knowledge of the instantaneous link conditions which can be mapped into a reliable transmission bit rate $R_i(t)$ at time $t$. If the available rate is below a given threshold $\bar{R_i}$, station $i$ gives up its transmission opportunity and frees up the channel for re-contention. Otherwise, the station transmits data for a fixed duration of time $\mathcal{T}$. \new{We illustrate the operation of DOS  in  Fig.~\ref{fig:system-model}.}

Our model, like that of \cite{itit,ton,twc,infocom,doc}, assumes that $R_i(t)$ remains constant for the duration of a data transmission and that different observations of $R_i(t)$ are independent.\footnote{The assumption that $R_i(t)$ remains constant during a transmission is a standard assumption for the block-fading channel in wireless communications \cite{mobicom02}, while the assumption of independent observations is justified in \cite{itit} through numerical calculations.} From \cite{itit}, we have that the optimal transmission policy is a threshold policy: given a threshold $\bar{R}_i$, station $i$ only transmits after a successful contention if $R_i(t) \geq \bar{R}_i$.

With the above model, stations' throughputs are a function of the access probabilities, $\mathbf{p}=\{p_1, \dots, p_N\}$, and the transmission rate thresholds, $\mathbf{\bar{R}}=\{\bar{R}_1, \dots, \bar{R}_N\}$.  \new{Given that a proportionally-fair allocation maximizes  $\sum_i \log r_i$\cite{Kelly}, where $r_i$ is the  throughput of station $i$, we define our problem as the following unconstrained optimization problem:}
\begin{equation}
 \new{\max_{{\bf \bar{R}},{\bf p}} \sum_i \log r_i}
\end{equation}

\subsection{Optimal $p_i$ configuration}
We start by computing the optimal configuration of the $\mathbf{p}$ parameters. The analysis to compute these parameters follows that of \cite{doc}, but it relies on different approximations, which are needed for the adaptive mechanism design that we present in \S\ref{sec-adaptive}.
%
To compute the optimal $p_i$ configuration, we start by expressing the throughput $r_i$ as a function of $\mathbf{p}$. Let $l_i$ be the average number of bits that station $i$ transmits upon a successful contention and $T_i$ be the average time it holds the channel. Then, the throughput of station $i$ is
\begin{equation}\label{eq-ri}
r_i = \frac{p_{s,i} l_i}{\sum_j{p_{s,j} T_j}+(1-p_s)\tau}\nonumber
\end{equation}
where $p_{s,i} = p_i \prod_{j \neq i}{(1-p_j)}$ is the probability that a mini slot contains a successful contention of station $i$ and $p_{s}$ is the probability that it contains any successful contention, $p_s = \sum_i{p_{s,i}}$.

Both $l_i$ and $T_i$ depend on $\bar{R}_i$. Upon a successful contention, a station holds the channel for a time $\mathcal{T}+\tau$ in case it transmits data and $\tau$ in case it gives up the transmission opportunity. In case the station uses the transmission opportunity, it transmits a number of bits given by $R_i(t)\mathcal{T}$. Thus, $T_i$ and $l_i$ can be computed as
$T_i = Prob(R_i(t) < \bar{R}_i)\tau + Prob(R_i(t) \geq \bar{R}_i)(\mathcal{T}+\tau)$
and
$l_i = \int_{\bar{R}_i}^{\infty}{r \mathcal{T} f_{R_i}(r) dr}$
where $f_{R_i}(r)$ is the pdf of $R_i(t)$.
Similarly as in \cite{doc}, let us define $w_i$ as 
\begin{equation}\label{eq-widef}
w_i = \frac{p_{s,i}}{p_{s,1}}
\end{equation}
where we take station 1 as reference. From the above equation, we have that $p_{s,i} = w_i p_s/\sum_j{w_j}$; substituting this into (\ref{eq-ri}) yields
\begin{equation}\label{eq-rirewritten}
r_i = \frac{w_i p_s l_i}{\sum_j{w_j p_s T_j}+\sum_j{w_j}(1-p_s)\tau}\nonumber
\end{equation}

In a slotted wireless system such as the one of this paper, the optimal access probabilities satisfy $\sum_i{p_i} = 1$ (see \cite{infocom05}), which results in the following optimal success probability $p_s$:
\begin{equation}\label{eq-1e}
p_s = \sum_i{p_i \prod_{j \neq i}{1-p_j}} \approx \sum_i{p_i} e^{-\sum_j{p_j}} = e^{-1}
\end{equation}

With the above, the problem of finding the $\mathbf{p}$ configuration that maximizes the proportionally fair rate allocation is thus equivalent to finding the $w_i$ values that maximize $\sum_i{\log(r_i)}$, given that $p_s = 1/e$. To obtain these $w_i$ values, we impose $\frac{\partial \sum_i{\log(r_i)}}{\partial w_i} = 0\nonumber$
which yields
$\frac{1}{w_i} - N \frac{p_s T_i + (1-p_s)\tau}{\sum_i{w_i p_s T_i}+\sum_j{w_j}(1-p_s)\tau} = 0$.
Combining this expression for $w_i$ and $w_j$, we obtain 
\begin{equation}
\frac{w_i}{w_j} = \frac{p_s T_j + (1-p_s)\tau}{p_s T_i + (1-p_s)\tau}\nonumber
\end{equation}

\new{Under the assumption of small $p_i$'s (the case of interest to exploit multiuser diversity with an opportunistic scheduler), $1-p_i \approx 1$, and thus $(1-p_i) /( 1 -p_j) \approx 1$, which leads to  $w_i/w_j \approx p_i/p_j$. Moreover, given that $p_s = 1/e$, the above can be rewritten as}
\begin{equation}\label{eq-fairness}
\frac{p_i}{p_j} = \frac{T_j + (e-1) \tau}{T_i + (e-1) \tau}
\end{equation}

Furthermore, the probability that a given mini slot is empty can be computed as follows,
\begin{equation}\label{eq-pe}
p_e = \prod_{i}{1-p_i} \approx e^{-\sum_i{p_i}} = e^{-1}
\end{equation}

We use a different approximation than \cite{doc}'s in order to remove any dependency \new{on} the number of stations, a result that we will exploit to design an algorithm that works well under non-saturation conditions too. Our simulation results show a very small performance impact for using this approximation instead, practically negligible for scenarios with $N>4$ stations.

With the above, we solve the optimization problem by finding the $\mathbf{p}$ values that solve the system of equations formed by (\ref{eq-fairness}) and (\ref{eq-pe}). The uniqueness of the solution of this system of equations can be proved as follows. Without loss of generality, let us take the access probability of station 1, $p_1$, as reference. From (\ref{eq-fairness}) we have that $p_i$ for $i \neq 1$ can be expressed as a continuous and monotone increasing function of $p_1$. Applying this to (\ref{eq-pe}), we have that the term ($\prod_{i}{1-p_i}$) is a continuous and monotone decreasing function of $p_1$ that starts at 1 and decreases to 0, while the right hand side is the constant value $1/e$. From this, it follows that there is a unique value of $p_1$ that satisfies this equation. Taking the resulting $p_1$ and computing $p_i \ \forall i \neq 1$ from (\ref{eq-fairness}), we have a solution to the system. Uniqueness of the solution is given by the fact that all relationships are bijective and any solution must satisfy (\ref{eq-pe}), which (as we have shown) has only one solution.

Hereafter, we denote the unique solution to the system of equations by $\mathbf{p^{*}}=\{p_1^{*}, \dots, p_N^{*}\}$. Note that determining $\mathbf{p^{*}}$ requires computing $T_i \ \forall i$, which depend on the optimal configuration of the thresholds $\mathbf{\bar{R}}$. In the following section we address the computation of the optimal $\mathbf{\bar{R}}$, which we denote by $\mathbf{\bar{R}^{*}}=\{\bar{R}_1^{*}, \dots, \bar{R}_N^{*}\}$.

\subsection{Optimal $\bar{R}_i$ configuration}\label{sec-optRi}

In order to obtain the optimal configuration of $\mathbf{\bar{R}}$, we need to find the transmission rate threshold of each station that, given the $\mathbf{p^{*}}$ computed above, optimizes the overall performance in terms of proportional fairness.

To this aim, we rely on Theorem 1 in \cite{doc} to find that the optimal configuration of the transmission rate thresholds is given by $\bar{R}_k^{*} = \bar{R}_k^{1}$, where $\bar{R}_k^{1}$ is the transmission rate threshold that optimizes the throughput of station $k$ when it is alone in the channel and contends with $p_k = 1/e$ (under the assumption that different channel observations are independent).
This is done in \cite{itit}, which uses \emph{optimal stopping theory} and finds that the optimal threshold can be obtained by solving the following fixed point equation:
\begin{equation}\label{eq-threshold}
E\left[R_i(t) - \bar{R}_i^{*}\right]^+ = \frac{\bar{R}_i^{*} \tau}{\mathcal{T}/e}
\end{equation}

Note that the above allows computing the threshold $\bar{R}_i^{*}$ of a station based on \emph{local information} only, as (\ref{eq-threshold}) does not depend on the other stations in the network and their radio conditions. In particular, the optimal threshold configuration is \emph{independent of the access probabilities $\mathbf{p}$}, which is crucial as it allows \new{us to independently design the mechanisms to adjust the configuration of $\mathbf{\bar{R}}$ and $\mathbf{p}$, respectively, as we  explain in the sequel.}

\section{ADOS Mechanism}\label{sec-adaptive}



In this section, we present the ADOS mechanism, which consists of two independent adaptive algorithms. The first algorithm determines the access probability used by a station, $p_i$, adjusting the value when the number of active stations in the network or their sending behavior change. The second algorithm determines the transmission rate threshold of a station, $\bar{R_i}$, adapting its value to the changing radio conditions of the station. Both algorithms together aim to drive the system to the optimal point of operation.
One of the key features of these algorithms is that they do not require to know the number of stations in the network, and they do not need to keep track of the behavior of the other stations or their channel conditions.

\subsection{Non-saturation conditions}\label{sec-nonsat}

The optimal configuration $\{\mathbf{p^*},\mathbf{\bar{R}^*}\}$ obtained in the previous section corresponds to the case where all stations are saturated. We next discuss how to consider the case when some of the stations are not saturated.
As we explained above, when all the stations are saturated, the optimal channel empty probability $p_e$ takes a constant value equal to $1/e$, independent of the number of stations. The first key approximation is to assume that this also holds when some of the stations are not saturated. 
\new{The rationale behind this assumption is that the impact of the aggregated load of several non-saturated stations is similar to the impact of a smaller number of saturated stations. 
Given that, as we show in \S\ref{sec-configuration}, the optimal $p_e$ does not depend on the number of stations in saturated conditions, we can assume that $p_e=1/e$ when there are non-saturated stations too.}

We have also seen in the previous section that, under saturation, the optimal transmission rate thresholds are constant values that only depend on the local radio conditions. The second key approximation is to assume that the optimal transmission rate thresholds take the same constant values under non-saturation. \new{The rationale is as follows.  Proposition 3.1 in \cite{itit} shows that, additionally to the local radio conditions, the optimal threshold also depends on the number of slots $K$ prior to a successful channel access. As the mechanism we describe below drives the system to a point of operation where $E[K]\!=\!1/p_s\!=\!e$ even if there are non-saturated stations, we can assume that the optimal threshold in this case  is the one given by (\ref{eq-threshold}) for saturated stations.

}

We next present the design of the algorithms to adjust $p_i$ and $\bar{R}_i$ that consider both saturation and non-saturation conditions following the two approximations exposed above.

%
%
%

\subsection{Adaptive algorithm for $p_i$}

Following the first approximation above, with ADOS each station implements an adaptive algorithm to configure the access probability $p_i$, with the goal of driving the channel empty probability to $1/e$, as given by (\ref{eq-1e}).

Driving the channel empty probability toward a constant optimum value fits well with the framework of \emph{classic control theory}. With these techniques, we measure the \emph{output signal} of the system and, by judiciously adjusting the \emph{control signal}, we aim at driving it to the \emph{reference signal}. A key advantage of using such techniques is that they provide the means for achieving a good tradeoff between the speed of reaction and stability while guaranteeing convergence, which is a major challenge when designing adaptive algorithms.

\begin{figure}

\centering
\includegraphics[width=\linewidth]{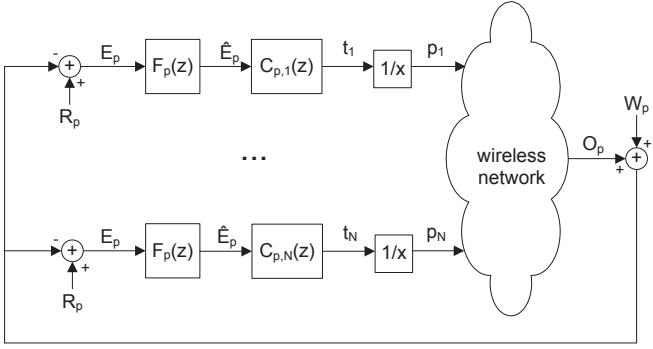}
\vspace{-5mm}
\caption{Adaptive algorithm for $p_i$.}
\vspace{-5mm}
\label{fig:system}
\end{figure}

Fig.~\ref{fig:system} depicts our algorithm to adjust $\mathbf{p}$, where each station computes the error signal $E_p$ by subtracting the output signal $O_p$ from the reference signal $R_p$ (the functions in the figure are given in the $z$ domain). The output signal $O_p$ is combined with a noise component $W_p$ of zero mean, modeling the randomness of the channel access algorithm. In order to eliminate this noise, we follow the design guidelines from \cite{birgitta} and introduce a low-pass filter $F_p(z)$. The filtered error signal $\hat{E}_p$ is then fed into the controller $C_{p,i}(z)$ of each station, which provides the control signal $t_i$, defined as the average time between two transmission of station $i$. Station $i$ then computes its access probability as $p_i = 1/t_i$. With the $p_i$ of each station, the wireless network provides the output signal $O_p$, which closes the loop.

In the above system, we need to design the reference and output signals $R_p$ and $O_p$, as well as the transfer functions of the low-pass filter and the controller, $F_p(z)$ and $C_{p,i}(z)$. We address next their design  with the goal of ensuring that the empty probability $p_e$ is driven to $1/e$.

In our system, time is divided into intervals such that the end of an interval corresponds to a transmission (either a success or a collision). Given that the target empty probability is equal to $1/e$, the target average number of empty mini slots between two transmissions (i.e., our reference signal) is equal to $R_p = 1/(e-1)$. In this way, after the $n$-th transmission, each station computes the output signal at interval $n$, denoted by $O_p(n)$, as the number of empty mini slots between the $(n-1)$-th and the $n$-th transmission. The error signal for the next interval is computed as
\begin{equation}
E_p(n+1) = R_p - O_p(n).
\end{equation}

With the above, if $p_e$ is too large then $O_p(n)$ will be larger than $R_p$ in average, yielding a negative error signal $E_p(n+1)$ that will decrease $t_i$ for the next interval, which will increase the transmission probability $p_i$ and therefore reduce $p_e$ (and vice-versa). This ensures that $p_e$ will be driven to the optimal value.

For the low-pass filter $F_p(z)$, we use a simple exponential smoothing algorithm of parameter $\alpha_p$ \cite{1204884}, given by the following expression in the time domain,
$\hat{E}_p(n) = \alpha_p E_p(n) + (1-\alpha_p)\hat{E}_p(n-1)\nonumber$,
which corresponds to the following transfer function in the $z$ domain:
$F_p(z) = \frac{\alpha_p}{1-(1-\alpha_p)z^{-1}}\nonumber$.
For the transfer function of the controllers $C_{p,i}(z)$, we use a simple controller from classical control theory, namely the Proportional Controller \cite{franklin}, which has already been used in a number of networking problems (e.g. \cite{boggia,tmc2}), i.e., 
$C_{p,i}(z) = K_{p,i}\nonumber$,
where $K_{p,i}$ is a per-station constant.

In addition to driving the empty probability to $1/e$, we also impose that the access probabilities satisfy (\ref{eq-fairness}).
Since we feed the same error into all stations, and the proportional controller simply multiplies this error by a constant to compute $p_i$, the following equation holds for all $i,j$:
\begin{equation}
\frac{p_i}{p_j} = \frac{K_{p,j}}{K_{p,i}}\nonumber
\end{equation}
Therefore, by simply setting $K_{p,i}$ as
$K_{p,i} = K_p\left(T_i +(e-1) \tau\right)\nonumber$,
we ensure that (\ref{eq-fairness}) is satisfied.

\subsection{Adaptive algorithm for $\bar{R}_i$}\label{sec-algorithm-ri}

Following the second approximation of \S\ref{sec-nonsat}, the adaptive algorithm of ADOS to adjust the threshold $\bar{R}_i$ aims to drive the threshold of all (saturated and non-saturated) stations to the optimal value given by (\ref{eq-threshold}). Note that (\ref{eq-threshold}) is equivalent to the following equation:
\begin{equation}\label{eq-avg}
E\left[(R_i(t) - \bar{R}_i^{*})^+ - \frac{\bar{R}_i^{*} \tau}{\mathcal{T}/e}\right] = 0
\end{equation}

\begin{figure}

\centerline{\includegraphics[width=\linewidth]{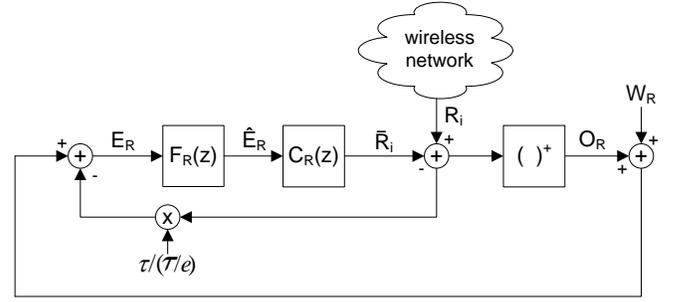}}
\vspace{-5mm}
\caption{Adaptive algorithm for $\bar{R}_i$.}%
\vspace{-5mm}
\label{fig:control}%
\end{figure}

In the following, we design an adaptive algorithm that drives $\bar{R}_i$ to the value given by the above equation. The algorithm is depicted in Fig.~\ref{fig:control}. Similarly to the adaptive algorithm for $p_i$, we base the algorithm design on control theory. The key difference between the two algorithms is that, since the optimal value of threshold of a station depends on local information only and hence does not depend on the threshold value of the other stations, we can consider each station separately (in contrast to Fig.~\ref{fig:system}).

In order to ensure that the configuration of $\bar{R}_i$ satisfies (\ref{eq-avg}), we design the output signal of the algorithm, $O_R$, equal to the term $(R_i - \bar{R}_i)^+$, and the reference signal, $R_R$, equal to the term $\bar{R}_i \tau/(\mathcal{T}/e)$. Thus, by driving the difference with these two terms (i.e., the error signal) to zero, we ensure that (\ref{eq-avg}) is satisfied.

Following the above, upon its $n^{th}$ successful contention, a station measures the channel transmission rate $R_i(n)$ and computes the output signal as
\begin{equation}
O_R(n)=
\begin{cases}
R_i(n) - \bar{R}_i(n), &  \text{if } R_i(n)>=\bar{R}_i(n) \\
0, & \text{otherwise}\nonumber
\end{cases}\nonumber
\end{equation}

From the above output signal, it then computes the error signal as
\begin{equation}
E_R(n+1) = O_R(n) - \frac{\bar{R}_i(n) \tau}{\mathcal{T}/e}\nonumber
\end{equation}

Due to the randomness of the radio signal, the output signal carries some noise $W_R$. In order to filter out this noise, we apply (like in the previous case) a low pass-filter $F_R(z)$ on the error signal, which yields
$\hat{E}_R(n) = \alpha_R E(n) + (1-\alpha_R)\hat{E}_R(n-1)\nonumber$.
Also like in the previous case, the error signal is introduced into a proportional controller,
$C_R(z) = K_R\nonumber$,
where $K_R$ is the constant of the controller.

The controller gives the threshold configuration $\bar{R}_i(n)$ as output. As mentioned above, by driving the error signal $\hat{E}_R(n)$ to 0, the controller ensures the threshold value satisfies (\ref{eq-avg}) and thus achieves the objective of adjusting the treshold to the optimal value $\bar{R}_i^{*}$ obtained in \S\ref{sec-configuration}.

\section{Control Theoretic Analysis}\label{sec-analysis}

With the above, we have all the components of the ADOS mechanism fully designed. The remaining challenge is the setting of its parameters, namely the parameters of the adaptive algorithm for $p_i$ ($K_{p}$ and $\alpha_{p}$) and the adaptive algorithm for $\bar{R_i}$ ($K_{{R}}$ and $\alpha_{{R}}$). In this section, we conduct a control theoretic analysis of the algorithms to find a suitable parameter setting.

As discussed in \S\ref{sec-configuration}, the setting of the optimal threshold $\bar{R}_i^{*}$ does not depend on the configuration of $\mathbf{p}$. Based on this, we analyze the closed-loop behavior of the two adaptive algorithms independently. For the adaptive algorithm to adjust $\bar{R}_i$, the behavior is independent of the $\mathbf{p}$ configuration. For the algorithm to adjust $p_i$, we consider that the values of $\mathbf{\bar{R}}$ are fixed, as their configuration depends only on the radio conditions, and analyze the convergence of $p_i$ to the optimal configuration corresponding to these $\mathbf{\bar{R}}$ values.


In the following, we first analyze the adaptive algorithm to adjust $p_i$ and then we analyze the one to adjust $\bar{R}_i$; these analyses provide good values for the parameters of the respective algorithms.

\subsection{Analysis of the algorithm for $p_i$}\label{sec-kp}

\begin{figure}

\centering
\includegraphics[width=\linewidth]{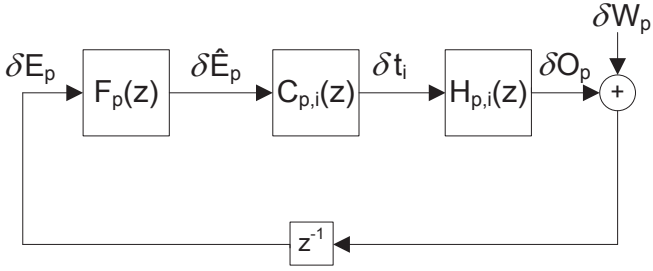}
\vspace{-5mm}
\caption{Closed-loop system of the adaptive algorithm for $p_i$.}
\vspace{-5mm}
\label{fig:station}
\end{figure}

We next conduct a control theoretic analysis of the closed-loop system of the algorithm for $p_i$ to find good values for the parameters $K_p$ and $\alpha_p$. Fig.~\ref{fig:station} depicts the closed-loop system for this algorithm. Note that the term $z^{-1}$ in the figure shows that the error signal $E$ at a given interval is computed with the output signal $O$ of the previous interval.

%
%
%

In order to analyze this system from a control theoretic standpoint, we need to characterize the transfer function $H_{p,i}$, which takes $t_i$ as input and gives $O_p$ as output. The following equation gives a nonlinear relationship between $O_p$ and $\{t_1,\ldots,t_N\}$:
\begin{equation}
O_p = \frac{1}{1-p_e} - 1\nonumber
\end{equation}
where $p_e = \prod_j \left( 1-1/t_j \right)$.

To express the above relationship as a transfer function, we linearize it when the system suffers small perturbations around its stable point of operation. Then, we study the linearized model and force that it is stable. Note that the stability of the linearized model guarantees that our system is locally stable.\footnote{\new{We assess stability from a control theory standpoint (a similar approach was used in~\cite{hollow} to analyze RED), in contrast to other analyses of schedulers such as \cite{weak-strong-stability} which look at the stability of the system queues from a queuing theory perspective.}}

We express the perturbations around the stable point of operation as follows:
\begin{equation}
t_i = t_i^{*} + \Delta t_i\nonumber
\end{equation}
where $t_i^{*} = 1/p_i^{*}$ is the stable point of operation of $t_i$, and $\Delta t_i$ are the perturbations around this point of operation.

With the above, the perturbations suffered by $O_p$ can be approximated by $\Delta O_p = \sum_{j}{\frac{\partial O_p}{\partial t_{j}} \Delta t_j}$ where
\begin{equation}
\frac{\partial O_p}{\partial t_{j}} = \frac{\partial O_p}{\partial p_{j}}\frac{\partial p_{j}}{\partial t_{j}} = \frac{p_e \, p_j^2}{(1-p_j)(1-p_e)^2}\nonumber.
\end{equation}

Given that $t_i/t_j = (T_i +(e-1) \tau)/(T_j +( e-1) \tau)$, the above can be rewritten as
\begin{equation}
\Delta O_p = \left(\sum_{j}{\frac{(T_j +(e-1) \tau)p_e \, p_j^2}{(T_i +(e-1) \tau)(1-p_j)(1-p_e)^2}}\right)\Delta t_i\nonumber
\end{equation}

With the above, we have characterized $H_{p,i}$:
\begin{equation}
H_{p,i} = \sum_{j}{\frac{(T_j +(e-1) \tau)p_e \, p_j^2}{(T_i +(e-1) \tau)(1-p_j)(1-p_e)^2}}\nonumber
\end{equation}

The closed-loop transfer function for station $i$ is then given by
\begin{equation}
T_{p,i}(z) = \frac{-z^{-1}C_{p,i}(z) F_p(z) H_{p,i}(z)}{1+z^{-1}C_{p,i}(z) F_p(z) H_{p,i}(z)}\nonumber
\end{equation}

Substituting the expressions for $F_p(z)$, $C_{p,i}(z)$ and $H_{p,i}(z)$ yields
\begin{equation}\label{eq-ti}
T_{p,i}(z) = \frac{-\alpha_p H_{p,i} K_{p,i}}{z - (1 - \alpha_p - \alpha_p K_{p,i} H_{p,i})}
\end{equation}

To guarantee stability, we need to ensure that the zero of the denominator of $T_{p,i}(z)$ falls inside the unit circle $|z| < 1$ \cite{astrom}, which implies
\begin{equation}
K_p < \frac{2-\alpha_p}{\alpha_p} \frac{1}{\sum_{j}{\frac{(T_j +(e-1) \tau)p_e \, p_j^2}{(1-p_j) (1-p_e)^2}}}\nonumber 
\end{equation}

The problem with the above upper bound is that it depends on the number of stations and their channel conditions. In order to assure stability, we need to obtain an upper bound that guarantees stability independent of these parameters. To do this, we observe that the right hand side of the above inequality takes a minimum value when $N=1$ and $T_1 = \tau + \mathcal{T}$. Therefore, by setting $K_p$ as follows, we guarantee that the above inequality will be met independent of the number of stations and their channel conditions:
\begin{equation}
K_p < K_p^{max} = \frac{2-\alpha_p}{\alpha_p\left(\mathcal{T} + e\tau\right)}\nonumber
\end{equation}


In order to set $K_p$ to a value that provides a good tradeoff between the speed of reaction to changes and stability, we follow the Ziegler-Nichols rules \cite{franklin}, which are widely used to configure proportional controllers. According to these rules, this parameter cannot be larger than one half of the maximum value that guarantees stability, which we denote by $K_p^{stability}$:
\begin{equation}\label{eq-kp}
K_p \leq K_p^{stability} = \frac{K_p^{max}}{2}
\end{equation}

In addition to the above, $K_p$ also needs to be set to eliminate the noise from the system. Noise is generated by the randomness of the output signal, which is given by the number of empty mini slots between two transmissions and hence follows a geometric random variable of factor $1-p_e = 1-1/e$. Hence, the noise at the input of the low-pass filter has a zero mean and a variance given by:
\begin{equation}
E[W_p^2] = \frac{p_e}{(1-p_e)^2}= \frac{1/e}{(1-1/e)^2}\nonumber
\end{equation}

The noise at the output of the controller can be obtained from the noise at the input of the low-pass filter with the following transfer function:
\begin{equation}
T_{W_p}(z) = \frac{-z^{-1}C_{p,i}(z)F_p(z)}{1+z^{-1}C_{p,i}(z)F_p(z)H_{p,i}(z)}\nonumber
\end{equation}

Substituting $C_{p,i}(z)$, $F_p(z)$ and $H_{p,i}(z)$ into the above yields
\begin{equation}
T_{W_p}(z) = \frac{-z^{-1}\alpha_p K_{p,i}}{1-z^{-1}(1-\alpha_p(1+K_{p,i}H_{p,i}))}\nonumber
\end{equation}

With the above transfer function, we can compute the variance of the noise at the output of the controller, denoted by $W_{p,c}$, as follows:
\begin{equation}
E[W_{p,c}^2] = \frac{\alpha_p^2 K_{p,i}^2}{1- (1-\alpha_p(1+K_{p,i}H_{p,i}))^2}\, E[W_p^2]\nonumber
\end{equation}

From the above equation, and taking into account from (\ref{eq-ti}) and (\ref{eq-kp}) that  $\alpha_p(1+K_{p,i}H_{p,i}) \leq 1 + \alpha_p/2$ we can obtain the following upper bound for $E[W_{p,c}^2]$:
\begin{equation}\label{wc}
E[W_{p,c}^2] \leq \frac{\alpha_p K_{p,i}}{(1-\alpha_p/2)H_{p,i}}\, E[W_p^2]\nonumber
\end{equation}

To limit the impact of the noise, we impose a gain factor of at least $G_p$ of the signal level at the output of the controller, $E[S_p^2]$, over the noise level at the same point, $E[W_{p,c}^2]$:
\begin{equation}
\frac{E[S_p^2]}{E[W_{p,c}^2]} \geq G_p\nonumber
\end{equation}

The signal at the output of the controller is equal to $t_i$, which yields $E[S_p^2] = t_i^2$. Combining this with the inequality of (\ref{wc}), we have that the following condition is sufficient to provide the desired gain:
\begin{equation}
\frac{t_i^2 (1-\alpha_p/2) H_{p,i}}{\alpha_p K_{p,i} E[W_p^2]} \geq G_p\nonumber
\end{equation}

Isolating $K_p$ from the above yields
\begin{equation}
K_p \leq \frac{t_i^2 (1-\alpha_p/2)}{G_p \alpha_p E[W_p^2]} \sum_{j}{\frac{(T_j +(e-1) \tau)p_e \, p_j^2}{(T_i +(e-1) \tau)^2(1-p_j)(1-p_e)^2}}\nonumber
\end{equation}
%
which is satisfied as long as the following condition holds,
\begin{equation}
K_p \leq \frac{1-\alpha_p/2}{G_p \alpha_p} \sum_{j}{\frac{T_j +(e-1) \tau}{(T_i +(e-1) \tau)^2}}\nonumber
\end{equation}

To find an upper bound that is independent of the number of stations and their conditions, we observe that the right hand side of the above inequality takes a minimum for $N = 1$ and $T_1 = \tau + \mathcal{T}$, which leads to the following upper bound, which we denote by $K_p^{noise}$,
\begin{equation}
K_p \leq K_p^{noise} = \frac{1-\alpha_p/2}{G_p \alpha_p\left(\mathcal{T} + e\tau\right)}\nonumber
\end{equation}

The analysis conducted in this section has given two upper bounds, $K_p^{stability}$ and $K_p^{noise}$, which guarantee that on the one hand the system is stable and on the other hand the noise level is not excessive. As these bounds depend on $\alpha_p$ and $G_p$, we also need to find a setting for these parameters. In order to provide a good level of protection against noise, $G_p$ needs to be sufficiently large. Additionally, in order to allow sufficiently large $K_{p,i}$ values, which is needed to avoid a large steady state error at the input of the controllers, $G_p\, \alpha_p$ needs to be sufficiently small. Following these considerations, we set $G_p = 10^2$ and $\alpha_p = 10^{-4}$. \new{With  $\alpha_p  = 10^{-4}$ we aim to mitigate the effect of the noise sufficiently, without compromising the speed of reaction to changes (i.e., in the order of magnitude of 1000 samples). With $G_p=10^2$ we set an upper bound to the noise power, i.e., we enforce a gain of the output signal of the controllers which is 100 times larger than the noise.}
With these $\alpha_p$ and $G_p$ values, we then configure $K_p$ as follows:

\begin{equation}
K_p = \min(K_p^{noise}, K_p^{stability})\nonumber
\end{equation}
which ensures that the two objectives concerning stability and noise are met.

\subsection{Analysis of the algorithm for $\bar{R}_i$}\label{sec-analysis-ri}

\begin{figure}

\centerline{\includegraphics[width=\linewidth]{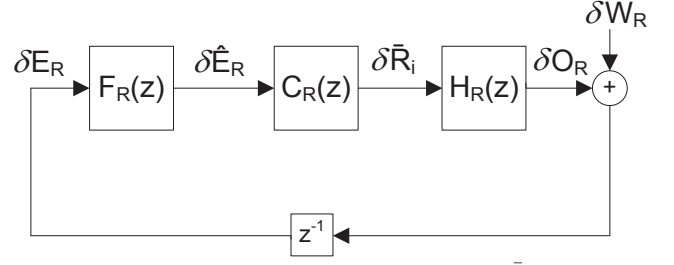}}
\vspace{-5mm}
\caption{Closed-loop system of the adaptive algorithm for $\bar{R}_i$.}%
\vspace{-5mm}
\label{fig:linear}%
\end{figure}

We next conduct a control theoretic analysis of the closed-loop system of the algorithm for $\bar{R}_i$, depicted in Fig.~\ref{fig:linear}. This analysis follows the same steps as the one above.

The perturbations around the point of equilibrium can be expressed as $\bar{R}_i = \bar{R}_i^* + \Delta \bar{R}_i$ and the perturbations suffered by $E_R$ can be approximated by $\Delta E_R = H_R \cdot \Delta \bar{R}_i$
where
\begin{align}
H_R = & \frac{\partial E_R}{\partial \bar{R}_i} = \frac{\partial}{\partial \bar{R}_i} \left((R_i - \bar{R}_i)^+ - \frac{\bar{R}_i \tau}{\mathcal{T}/e}\right) = \frac{\partial (R_i - \bar{R}_i)^+}{\partial \bar{R}_i} \nonumber - \frac{\tau}{\mathcal{T}/e}\nonumber
\end{align}


To compute $\partial (R_i - \bar{R}_i)^+/\partial \bar{R}_i$, we note that $(R_i - \bar{R}_i)^+$ expresses an average value, as the variations around this average value are captured by another component, namely the noise $W_R$. For the calculation of the average, we take all possible $R_i$ values weighted by $R_i$'s pdf, $f_{R_i}(r)$, which yields
\begin{align}
\frac{\partial (R_i - \bar{R}_i)^+}{\partial \bar{R}_i} &= \frac{\partial}{\partial \bar{R}_i}\int_{\bar{R}_i}^{\infty}{(r-\bar{R}_i)f_{R_i}(r) dr}  = - \int_{\bar{R}_i}^{\infty}{f_{R_i}(r) dr}\nonumber
\end{align}

With the above, $H_R$ can be expressed as $H_R = - H_{R,1} - H_{R,2}$, where $H_{R,1} = e\tau/\mathcal{T}$ and $0 \leq H_{R,2} \leq 1$.

The closed-loop transfer function of the system is given by
\begin{equation}
T_R(z) = \frac{C_R(z) F_R(z) H_R(z)}{1-z^{-1}C_R(z) F_R(z) H_R(z)}\nonumber
\end{equation}
where
\begin{equation}
F_R(z) = \frac{\alpha_R}{1-(1-\alpha_R)z^{-1}},~~~  C_R(z)=K_R.\nonumber
\end{equation}

Substituting the expressions for $F_R(z)$, $C_R(z)$ and $H_R(z)$ yields
\begin{equation}
T_R(z)  = \frac{-\alpha_R K_R (H_{R,1}+H_{R,2})}{1 - z^{-1}(1-\alpha_R - K_R \alpha_R (H_{R,1}+H_{R,2}))}\nonumber
\end{equation}

To guarantee stability, we need to ensure that the zero of the denominator of $T_R(z)$ falls inside the unit circle $|z| < 1$, which implies
\begin{equation}
K_{R} < \frac{2-\alpha_{R}}{\alpha_{R} (H_{R,1}+H_{R,2})}\nonumber
\end{equation}

In order to find a sufficient condition that holds for all cases, we consider the worst case $H_{R,2} = 1$, which leads to 
\begin{equation}
K_R < \frac{2-\alpha_R}{\alpha_R (1+e\tau/\mathcal{T})}\nonumber
\end{equation}

According to Ziegler-Nichols rules, to guarantee stability we take a $K_R$ value equal to half of the above value,
\begin{equation}
K_R^{stability} = \frac{2-\alpha_R}{2 \alpha_R (1+e\tau/\mathcal{T})}\nonumber
\end{equation}

The noise introduced into the system, $W_R$, is given by the randomness in the transmission rate values $R_i$. If we assume that the available transmission rate for a given SNR is given by the Shannon channel capacity, then $R_i = C \log(1+\rho |h|^2)$, where $C$ is a constant parameter, $\rho |h|^2$ is the SNR and $h$ is the normalized random gain of the channel ($E[h] = 1$). Note that the values of $R_i$ below $\bar{R}_i$ are eliminated from the system by the module that performs the operation $(R_i - \bar{R}_i)^+$, which reduces the noise in the system. In what follows, we do not consider this effect in order to obtain an upper bound on the noise, which provides a worst case analysis.

If we represent the SNR as the sum of its average value ($\rho$) plus some noise of zero mean (which we denote by $W_h$), then we can express the transmission rates $R_i$ as $R_i = C \log (1+ \rho + W_h)$
which we can approximate at the stable point of operation ($W_h = 0$) by
\begin{equation}
R_i \approx C \log (1+ \rho) +  W_h\frac{\partial R_i}{\partial W_h}\bigg|_{W_h = 0}\nonumber
\end{equation}

Since the noise introduced into the system is given by the variations of $R_i$ around its average value, from the above we have that we can approximate $W_R$ by
\begin{equation}
W_R \approx  W_h\frac{\partial R_i}{\partial W_h}\bigg|_{W_h = 0} = \frac{C}{1 + \rho} W_h\nonumber
\end{equation}

With the above approximation, we can compute the variance of $W_R$ as follows,
\begin{equation}
E[W_R^2] = \frac{C^2}{(1+\rho)^2}E[W_h^2]\nonumber
\end{equation}

If we assume that the channel follows a Rayleigh fading model, then $\rho |h|^2$ corresponds to an exponential random variable of rate $\rho^{-1}$. With this, we have that $E[W_h^2] = \rho^2$, which yields $E[W_R^2] = \frac{C^2\rho^2}{(1+\rho)^2}$.

If we denote the noise at the output of the controller by $W_{R,c}$, we have
\begin{equation}
W_{R,c}(z) = \frac{F_R(z)C_R(z)}{1-z^{-1}F_R(z)C_R(z)H_R(z)} W_R(z)\nonumber
\end{equation}
from which
\begin{equation}
W_{R_c}(z) = \frac{\alpha_R K_R}{1 - z^{-1}(1-\alpha_R - K_R \alpha_R (H_{R,1}+H_{R,2}))} W_R(z)\nonumber
\end{equation}

From the above, the variance of the noise at the output of the controller can be computed as
\begin{equation}
E[W_{R,c}^2] = \frac{(\alpha_R K_R)^2}{1 - (1-\alpha_R(1+K_R(H_{R,1}+H_{R,2})))^2} E [W_R^2]\nonumber
\end{equation}

Given that $K_R \leq K_R^{stability}$, we can obtain the following upper bound on $E[W_{R,c}^2]$:
\begin{equation}\label{eq-wrc}
E[W_{R,c}^2] \leq \frac{\alpha_R K_R}{(H_{R,1}+H_{R,2})(1 - \alpha_R/2)} E [W_R^2]
\end{equation}

In order to guarantee a gain of $G_R$ of the signal over the noise at the output of the controller, we impose
\begin{equation}\label{eq-gr}
\frac{E[S_R^2]}{E[W_{R,c}^2]} \geq G_R
\end{equation}
where the signal is the threshold $\bar{R}_i$, which we approximate by the average transmission rate, $C \log(1+\rho)$. With this and the upper bound of (\ref{eq-wrc}) for $E[W_{R,c}^2]$, we can obtain the following sufficient condition to guarantee (\ref{eq-gr}):
\begin{equation}
\left(\frac{\log(1+ \rho) (1+\rho)}{\rho}\right)^2 \frac{(H_{R,1}+H_{R,2})(1-\alpha_R/2)}{\alpha_R K_R} \geq G_R\nonumber
\end{equation}

Isolating $K_R$ from the above yields
\begin{equation}
K_R \leq \left(\frac{\log(1+ \rho) (1+\rho)}{\rho}\right)^2 \frac{(H_{R,1}+H_{R,2})(1-\alpha_R/2)}{\alpha_R G_R}\nonumber
\end{equation}

In order to find a value of $K_R$ that ensures the desired gain for all scenarios, we chose the $\rho$ value that minimizes the right hand side of the above equation and take the worst case value for $H_{R,1}$, which leads to the following upper bound on $K_R$, which we denote by $K_R^{noise}$,
\begin{equation}
K_R \leq K_R^{noise} = \frac{e \tau (1-\alpha_R/2)}{\mathcal{T} \alpha_R G_R} \nonumber
\end{equation}

\new{Following the rationale of} \S\ref{sec-kp}, we set $G_R = 10^2$ and $\alpha_R = 10^{-4}$ and choose $K_R = \min(K_R^{noise},K_R^{stability})$, which ensures that the two goals in terms of noise and stability are met.

\section{Performance Evaluation}\label{sec-performance}

In this section, we present a performance evaluation of ADOS by means of simulations. Unless otherwise stated, we assume that different observations of the channel conditions are independent and that the available transmission rate for a given SNR is given by the Shannon channel capacity: $R(h) = B \log_2(1+\rho |h|^2) \ \ \textnormal{bits/s}$, where $B$ is the channel bandwidth in Hz, $\rho$ is the normalized average SNR and $h$ is the random gain of Rayleigh fading.
\new{Unless otherwise stated, we set $\mathcal{T}/\tau=10$ and $\rho=1$, i.e., the same values used in \cite{itit}, for comparison purposes. We also set $B=10^7$ and run enough simulations to obtain 95\% confidence intervals below 1\%.} 

\subsection{\new{Homogeneous scenario}}\label{sec-perf-first}

\subsubsection{Saturated stations}
We start by considering a homogeneous scenario where all stations are saturated and have the same normalized average SNR ($\rho_i= 1 \ \forall i$). We compare the performance of ADOS to the following approaches:
\begin{itemize}
\item[($i$)] The static optimal configuration obtained from performing an exhaustive search over the $\{p_i,\bar{R_i}\}$ space and choosing the best configuration (`\emph{static configuration}').
\item[($ii$)] An approach that\new{, although it probes the channel too (to avoid long collisions), it  never skips a transmission opportunity regardless of the estimated link quality} (`\emph{non-opportunistic}').
\item[($iii$)]\new{A CSMA/CA protocol which does not skip any transmission opportunity but it does not probe the channel so collisions last for the duration of a frame.}
\item[($iv$)] The team game approach proposed in \cite{itit} (TDOS). This approach requires that each station knows the channel state of all the stations in the network, and hence incurs substantial signaling overhead.  \new{In the simulations we assume that this overhead is non-existent}.
\item[($v$)] The non-cooperative approach proposed in \cite{itit} (NDOS). This approach, like ours, only requires information that can be observed locally, and hence does not involve any signaling.\footnote{Since \cite{itit} only optimizes the transmission rate thresholds but not the access probabilities, we take the ${p_i}$'s used in the simulations of \cite{itit} for TDOS and NDOS. For `\emph{non-opportunistic}', we choose the access probabilities that maximize performance.}
\end{itemize}

Fig.~\ref{fig:homogeneous} shows the total throughput as a function of the number of stations in the network. The figure confirms that ADOS is effective in driving the system to the optimal point of operation, providing the same throughput as the benchmark given by the `\emph{static configuration}'. The TDOS and NDOS approaches provide lower throughput as they only optimize the transmission rate thresholds; among them, NDOS performs substantially worse as it has less information. Finally, the `\emph{non-opportunistic}' approach provides the lowest throughput due to the lack of opportunistic scheduling. In conclusion, the proposed ADOS mechanism provides optimal throughput performance, outperforming the other approaches.


%

\subsubsection{Non-saturated stations}

\begin{figure}[tb!]        

\centering
                \centering
	    \includegraphics[width=\linewidth]{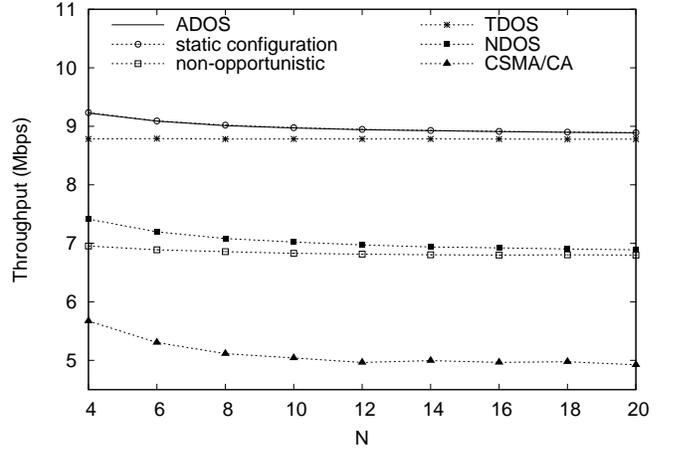}
	    \label{fig:homogeneous}

  \caption{\new{Homogeneous scenario with $N$ saturated stations.}}
  
  \label{fig:homogeneous}
\end{figure}

\begin{figure}[tb!]        

\centering
             \subfloat[]{
	    \includegraphics[width=0.5\linewidth ]{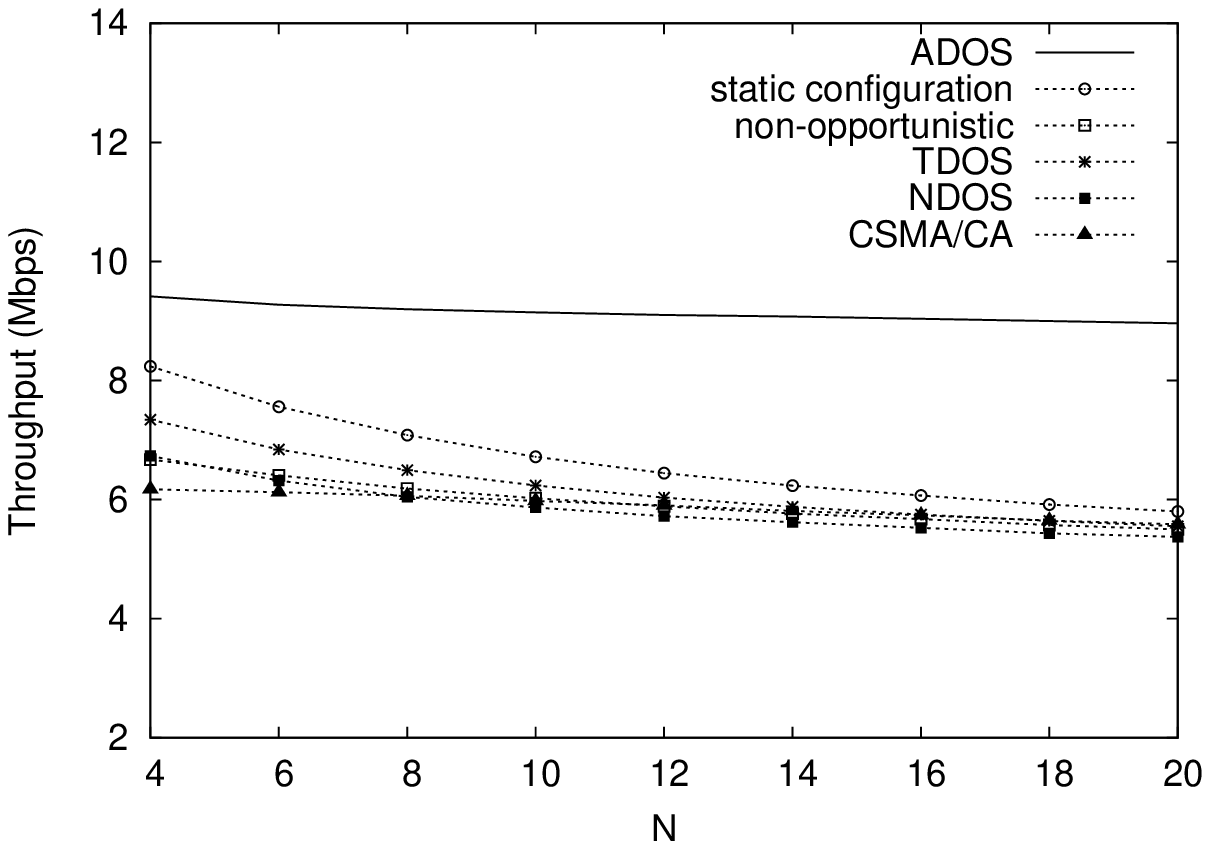}
	    \label{fig:non_saturation}
        }%
        \subfloat[]{
                \centering
	    \includegraphics[width=0.5\linewidth]{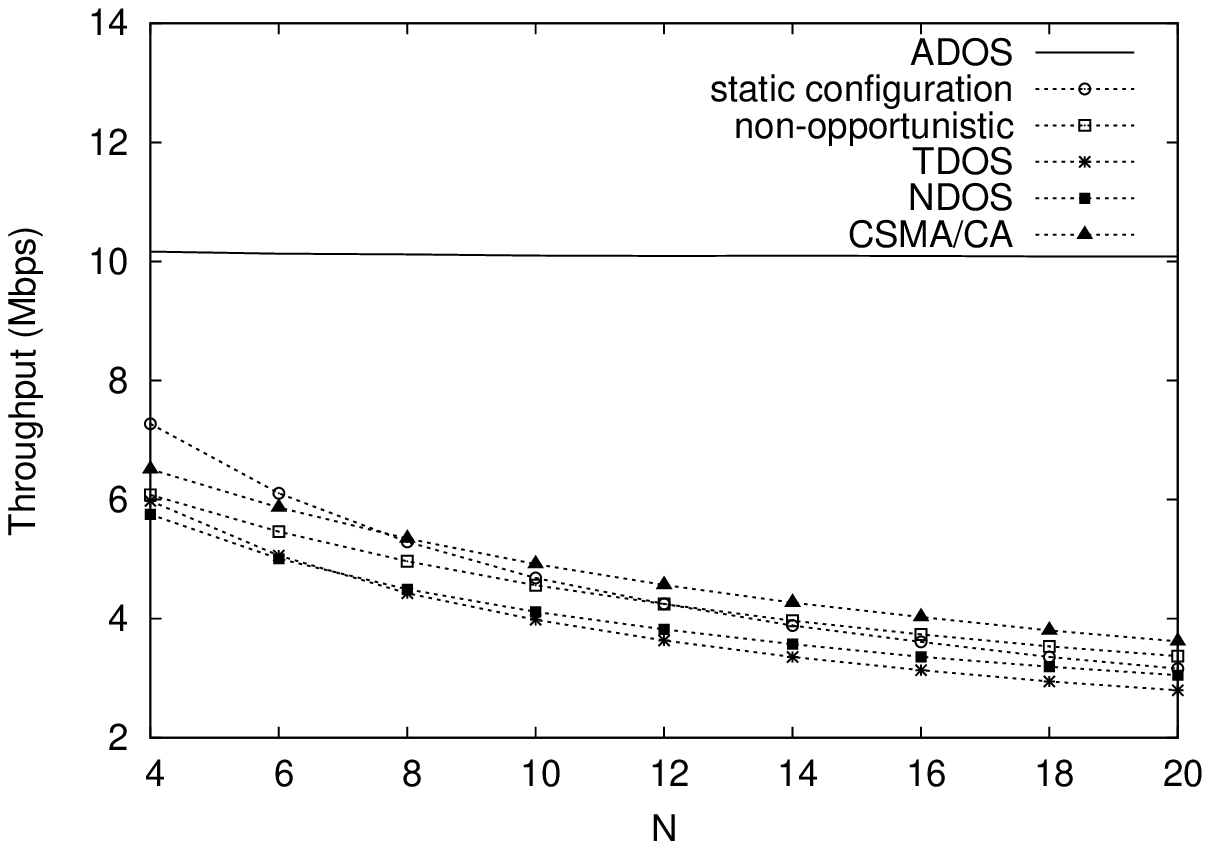}
	    \label{fig:non_saturation_low}
        }
  
  \caption{\new{Homogeneous scenario with (a) $N-1$ with medium load, and (b) $N-1$  with low load.}}
  
  \label{fig:nonsaturation}
\end{figure}

 We now assess the performance in the presence of non-saturation stations (that do not always have data ready for transmission). 
We first consider a scenario with homogeneous radio conditions ($\rho_i= 1 \ \forall i$) with one saturated station and $N-1$ non-saturated stations. Figs.~\ref{fig:non_saturation} and \ref{fig:non_saturation_low} illustrate the total throughput of the  network as a function of the number of stations, when the non-saturated stations transmit  at one half and one tenth of their saturation throughput (i.e., the throughput the would obtain if they were saturated). We observe that ADOS significantly outperforms all other approaches and that this effect becomes more accentuated as the throughput of the non-saturated stations decreases. The reason is that the other approaches assume that all stations are always saturated, and thus the access probabilities they use become overly conservative for the non-saturated case.

\subsection{Heterogeneous scenario}\label{sec-perf-second}

\begin{figure}[t!]

        \centering
             \subfloat[]{
	    \includegraphics[width=0.5\linewidth ]{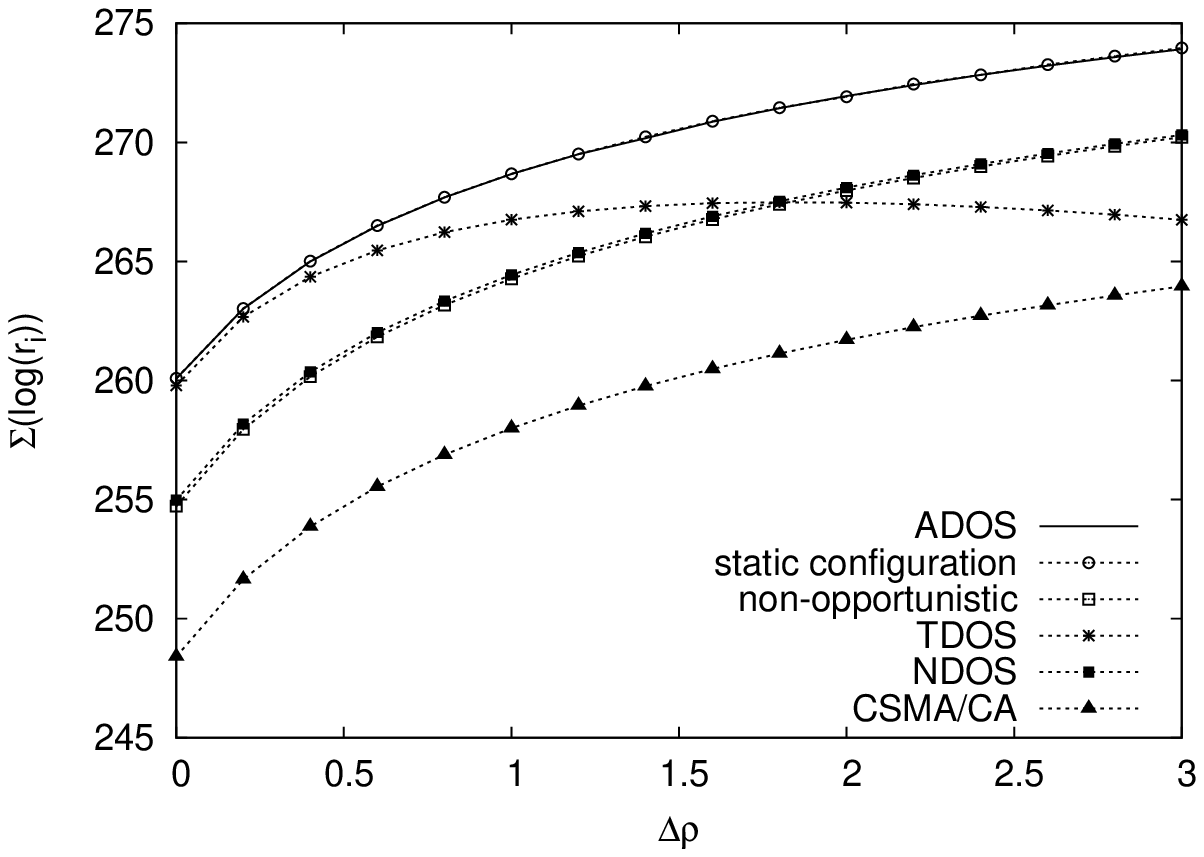}
	    \label{fig:heterogeneous}
        }%
        \subfloat[]{
                \centering
	    \includegraphics[width=0.5\linewidth]{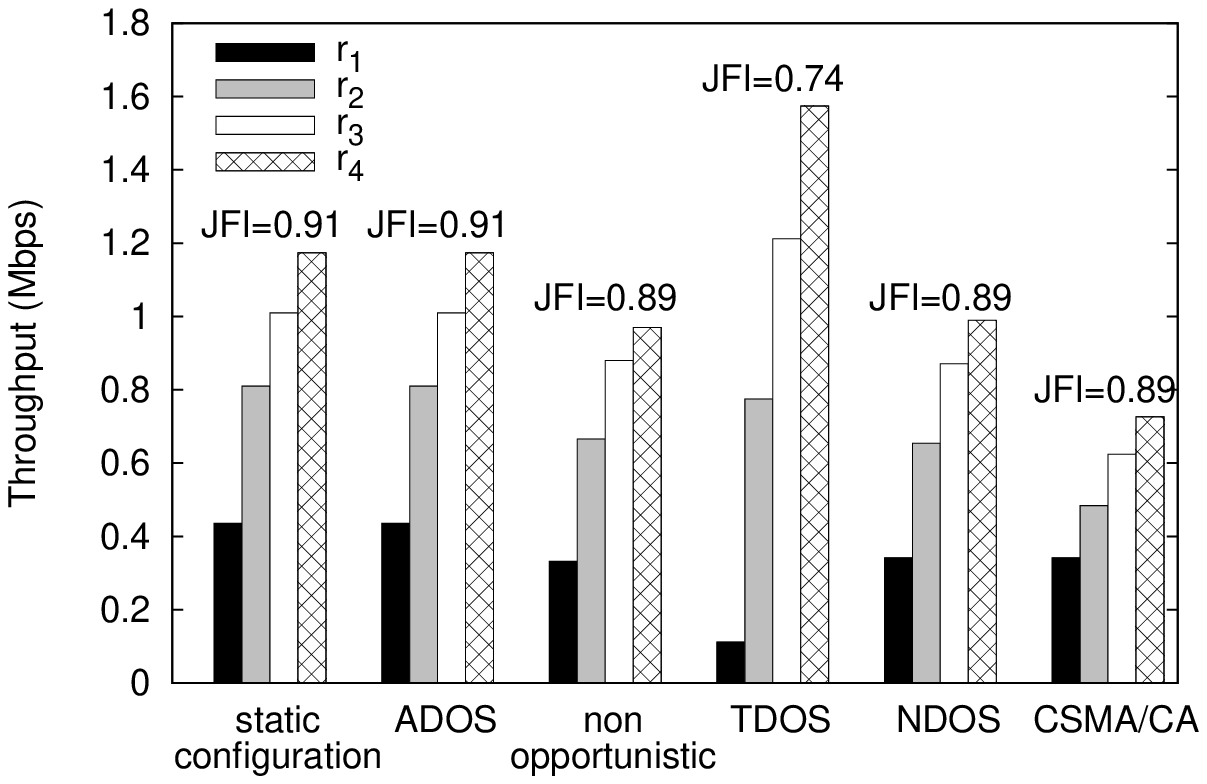}
	    \label{fig:heterogeneous2}
        }
  
  \caption{\new{Heterogeneous scenario. (a) $\sum_i{\log(r_i)}$ as a function of $\Delta \rho$. (b) Throughput of a station of each group for $\Delta \rho = 2$.}}
  
  \label{fig:heterogeneous_all}
\end{figure}

In the case of heterogeneous channel conditions, performance does not only depend on the total throughput but also on the way this throughput is shared among the stations. To analyze performance in this scenario, we consider $N=20$ saturated stations divided into four groups according to their channel conditions. The normalized SNR of the stations from group $i$ is given by $\rho_i=1+(i-1)\Delta \rho$, with $i \in \{1,2,3,4\}$.  Fig.~\ref{fig:heterogeneous} shows $\sum_i{\log(r_i)}$, the figure of merit for proportional fairness, as a function of $\Delta \rho$. 
 We observe that ADOS performs at the same level as the benchmark given by the `\emph{static configuration}', while the other approaches provide a substantially lower performance. TDOS exhibits an increasing degree of unfairness as $\Delta \rho$ grows that harms its performance in terms of proportional fairness. NDOS, in contrast to TDOS, does not show this behavior: with NDOS, each station sets its threshold based on its local radio conditions and therefore the fact that other stations have better radio conditions does not impact fairness. The price that NDOS pays for this non-cooperative behavior, however, is that the overall throughput performance is substantially degraded for all $\Delta \rho$ values. The `\emph{non-opportunistic}' approach \new{and CSMA/CA} also provide poor performance, similar to NDOS.

In order to gain additional insight into the throughput distribution with heterogeneous radio conditions, Fig.~\ref{fig:heterogeneous2} depicts the throughput obtained by a station of each group with the different approaches, along with the Jain's fairness index (JFI) \cite{jain} of each distribution. The results confirm that TDOS suffers from high unfairness with heterogeneous radio conditions, since with this approach the stations with worst radio conditions ($r_1$) are almost starved while the stations with best radio conditions ($r_4$) obtain a very large throughput. In contrast, the TDOS, `\emph{non-opportunistic}' \new{and CSMA/CA} approaches do not suffer from unfairness but provide significantly smaller throughputs than ADOS. We conclude that ADOS substantially outperforms all other approaches with heterogeneous radio conditions.



\subsection{\new{Performance under realistic models}}

\subsubsection{Impact of channel coherence time}

Our channel model is based on the assumption that different observations of the channel conditions are independent. In order to understand the impact of this assumption, we repeated the experiment of \new{Fig.~\ref{fig:heterogeneous}} using \emph{Jakes' channel model} \cite{jakes} to obtain channel conditions that are correlated over time. The results, for a Doppler frequency of $f_D = 2 \pi /100 \tau$ \new{(which roughly corresponds to 100 Km/h at 2.4 GHz)}, are given in Fig.~\ref{fig:channel model} where ADOS outperforms all the others. We also observe that the performance is \new{slightly} lower than that of \new{Fig.~\ref{fig:heterogeneous}}. This is due to the fact that when the channel is bad, a station does not transmit after a successful contention, and therefore it takes a shorter time until it successfully contends again. Thus, a station accesses the channel more often when the channel is bad than when it is good, which introduces a bias that reduces throughput. 


\subsubsection{Discrete set of transmission rates}\label{sec-discrete}

\begin{figure*}[t!]        

\centering
             \subfloat[]{
	    \includegraphics[width=0.33\linewidth ]{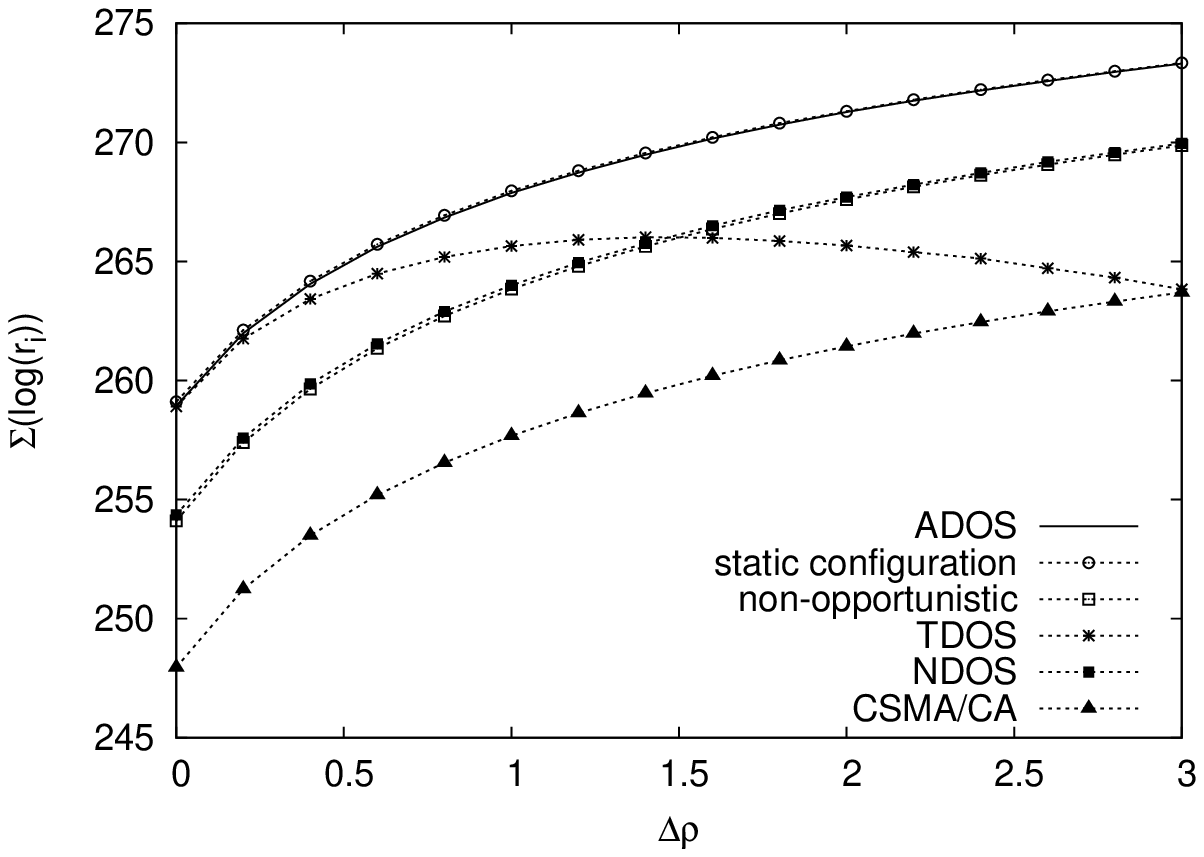}
	    \label{fig:channel model}
        }%
        \subfloat[]{
                \centering
	    \includegraphics[width=0.33\linewidth]{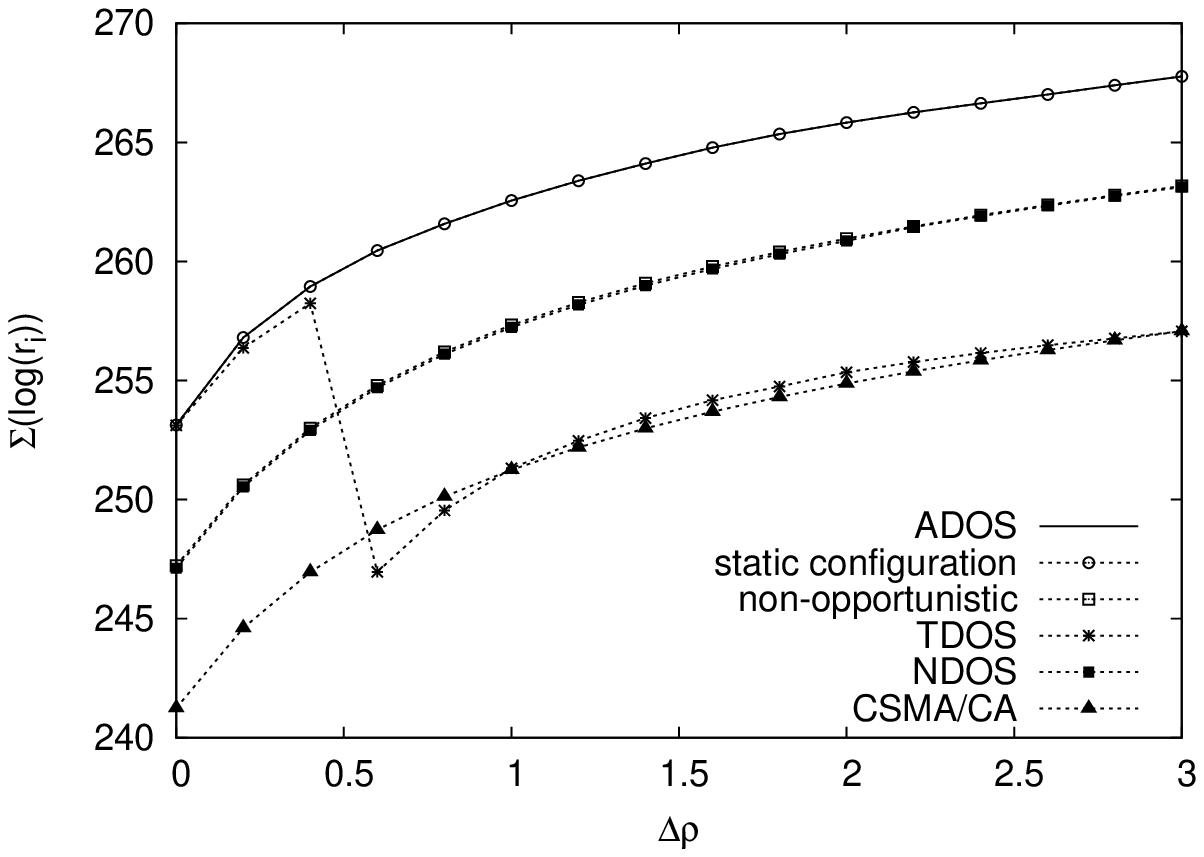}
	    \label{fig:discrete_rates}
        }
        \subfloat[]{
                \centering
	    \includegraphics[width=0.33\linewidth]{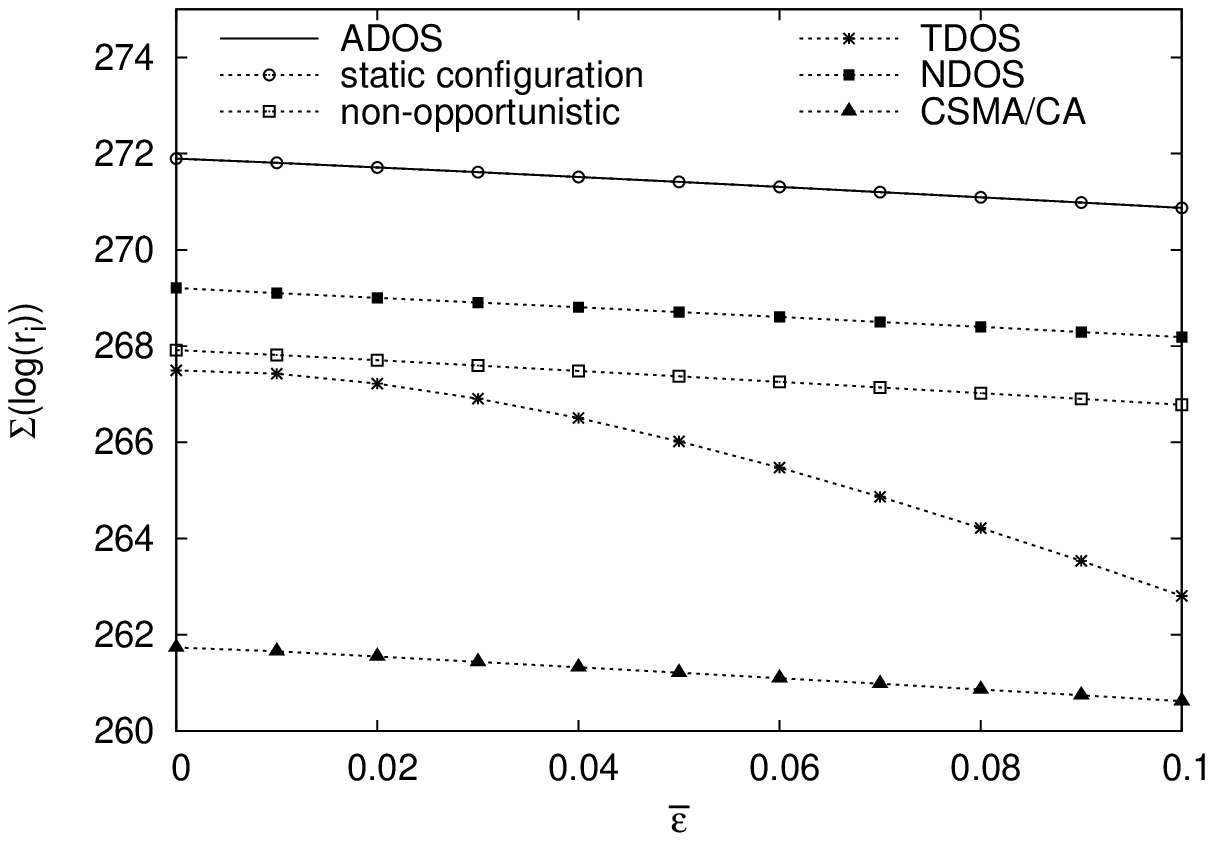}
	    \label{fig:imperfect}
        }        
  
  \caption{\new{Heterogeneous scenario. (a) \emph{Jakes'} channel model. (b) Discrete set of rates. (c) Imperfect channel quality estimation. }}
  
  \label{fig:other_channel}
\end{figure*}

While all previous experiments assumed continuous rates, the design of ADOS do not rely on any assumption on the mapping of SNR to transmission rates, and therefore any mapping function (continuous or discrete) can be used. We consider the case of a wireless system in which the only transmission rates available are $\{1, 2, 5.5, 12, 24, 48, 54\}$ Mbps. For a given SNR, we choose the largest available transmission rate that is smaller than the one given by Shannon channel capacity model.  Fig.~\ref{fig:discrete_rates} shows the result of repeating the experiment of \new{Fig.~\ref{fig:heterogeneous}} with this discrete set of transmission rates. The results confirm that ADOS outperforms the other approaches \new{with different mapping functions}.

\subsubsection{Imperfect channel estimation}\label{sec-imperfect}

\new{
Our design assumes that the channel state is perfectly known to the transmitter. However, real estimators often have to deal with noisy observations and produce inaccurate results, which may worsen performance or even cause outage in the communication.
Yet, according to \cite{twc}, the optimal threshold still has a threshold structure under these conditions. To assess the performance in the presence of estimation errors, we model the measured  SNR as $\rho_{meas} = \rho(1-\varepsilon)$, where $\varepsilon$ is the random estimation error with average $\bar{\varepsilon}$, and, following the scheme proposed in \cite{twc}, we select a linear  function to back off from the estimated bit rate which is equal to $\bar{\varepsilon}$. We evaluate the same heterogeneous scenario as before for  $\Delta \rho = 2$, and plot in Fig.~\ref{fig:imperfect} the performance as a function of $\bar{\varepsilon}$ for all the schemes under evaluation, revealing that ADOS also outperforms all the others in this case. 
}

\subsection{\new{Validation of the configuration proposed for ADOS}}\label{sec-validation-conf}\label{sec-perf-last}

\new{
The analysis in \S\ref{sec-analysis} derives the guidelines to configure the parameters of ADOS ($\{K_p,\alpha_p\}$  and $\{K_R,\alpha_R\}$) in order to guarantee a good behavior over time (stability and convergence speed). We next validate such guidelines in contrast to other settings that deviate from them.
}



\subsubsection{\new{Static conditions}}

\begin{figure}[t!]    

\centering
             \subfloat[]{
	    \includegraphics[width=0.5\linewidth]{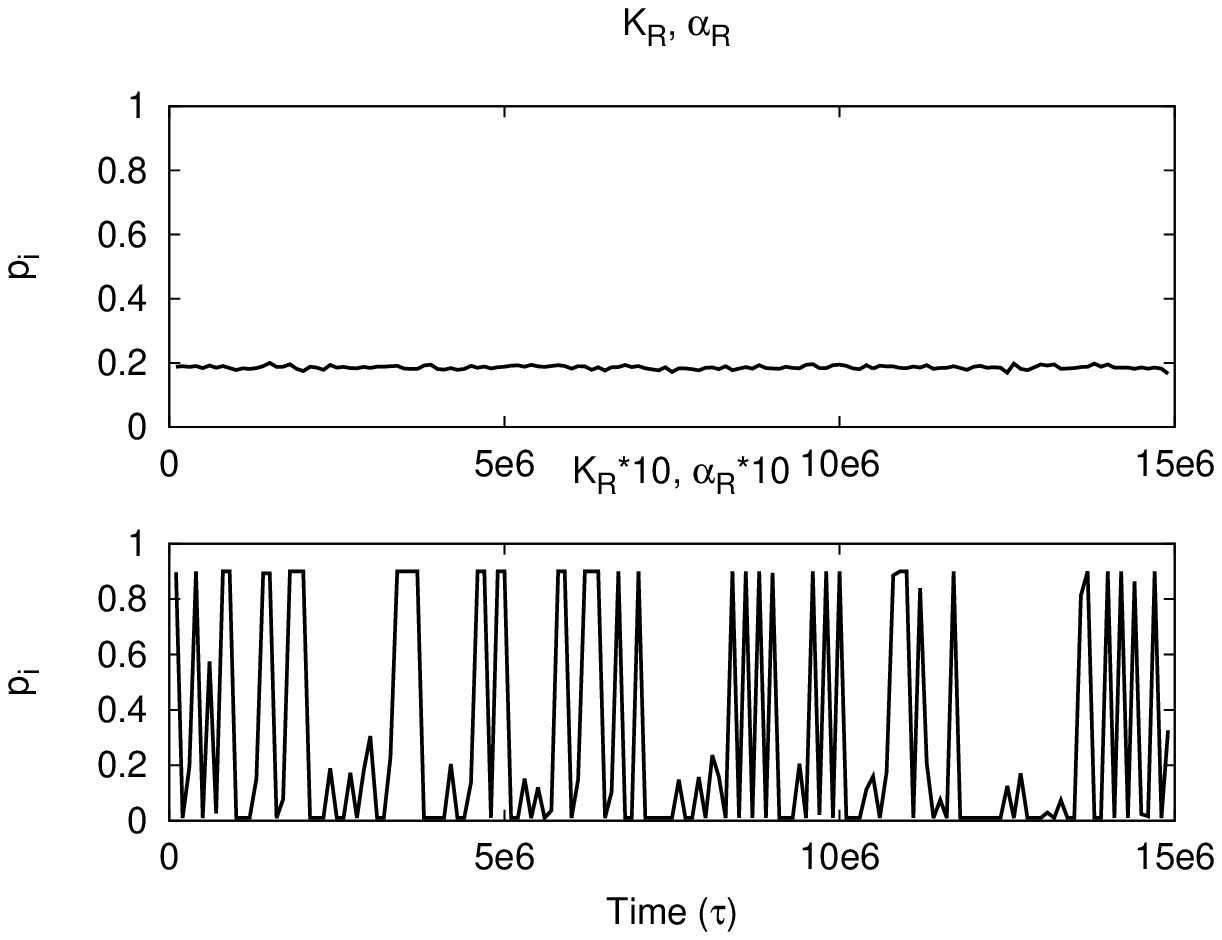}
	    \label{fig:stability_pi}
        }%
        \subfloat[]{
                \centering
	    \includegraphics[width=0.5\linewidth]{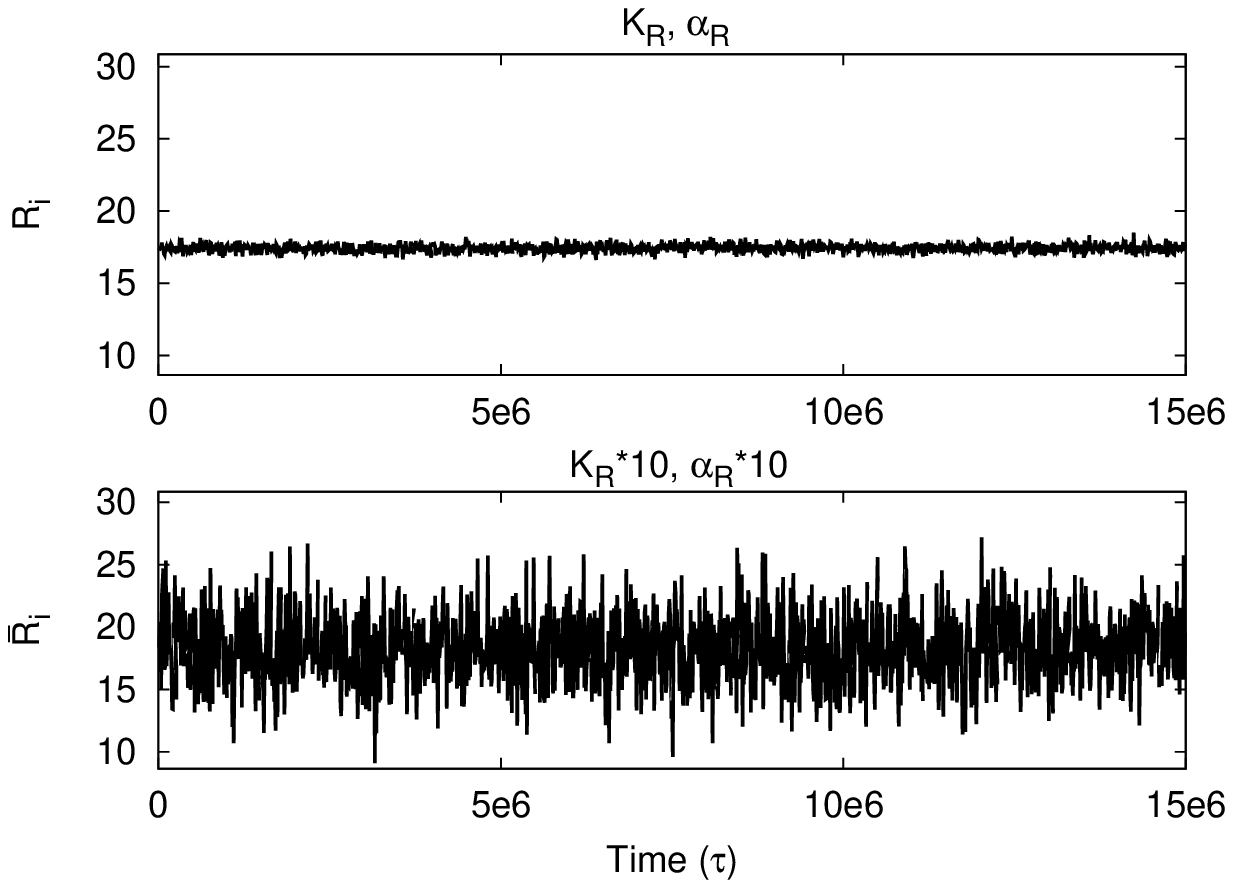}
	    \label{fig:stability_Ri}
        }
           
  \caption{\new{Validation in static conditions of the proposed configuration (a) $p_i$ and (b) $R_i^*$.}}
  
  \label{fig:stability}
\end{figure}

To verify stable behavior \new{in a static environment}, we first \new{observe} the evolution over time of the access probability $p_i$ of a station for the proposed $\{K_p,\alpha_p\}$ setting and for a configuration of these parameters 10 times larger, in a homogeneous scenario with $N=5$ saturated stations and $\rho = 4$. Fig.~\ref{fig:stability_pi} shows the evolution of $p_i$ for both cases, sampled over $10^5 \tau$ intervals. We observe from the figure that with the proposed setting (labeled ``$K_p,\alpha_p$''), $p_i$ shows minor deviations around its average value, while for a larger setting (labeled ``$K_p*10,\alpha_p*10$''), it shows unstable behavior with drastic oscillations.

Similarly, we also \new{observe} the evolution over time of the threshold $\bar{R}_i$ of a station for the proposed $\{K_R,\alpha_R\}$ setting and for a configuration of these parameters 10 times larger in the same scenario. The results, depicted in Fig.~\ref{fig:stability_Ri} confirm that the proposed setting for these parameters is stable while a larger setting is highly unstable. We conclude from these results that the analysis conducted in \S\ref{sec-analysis} is effective in guaranteeing stability.

%

\subsubsection{\new{Changing number of stations}}

We next investigate the speed with which the system reacts to changes in the number of stations of the  network, which triggers the adjustment of the access probabilities $p_i$. To this aim, we consider a  network with initially 5 stations, where 5 additional stations join the network after a time $5\!\cdot\!10^6 \tau$. 
Fig.~\ref{fig:reaction_pi} shows the evolution of the access probability of one of the initial stations sampled over $10^5 \tau$ intervals. We observe from the figure that with our setting (labeled ``$K_p,\alpha_p$''), the system quickly adapts the $p_i$ of the station to the new value. In contrast, for a setting of these parameters 10 times smaller (labeled ``$K_p/10,\alpha_p/10$''), the reaction is very slow and the system only converges after $5\!\cdot\!10^6 \tau$.



%

\subsubsection{\new{Changing radio conditions}}

\begin{figure*}[t!]    

\centering
             \subfloat[]{
	    \includegraphics[width=0.33\linewidth]{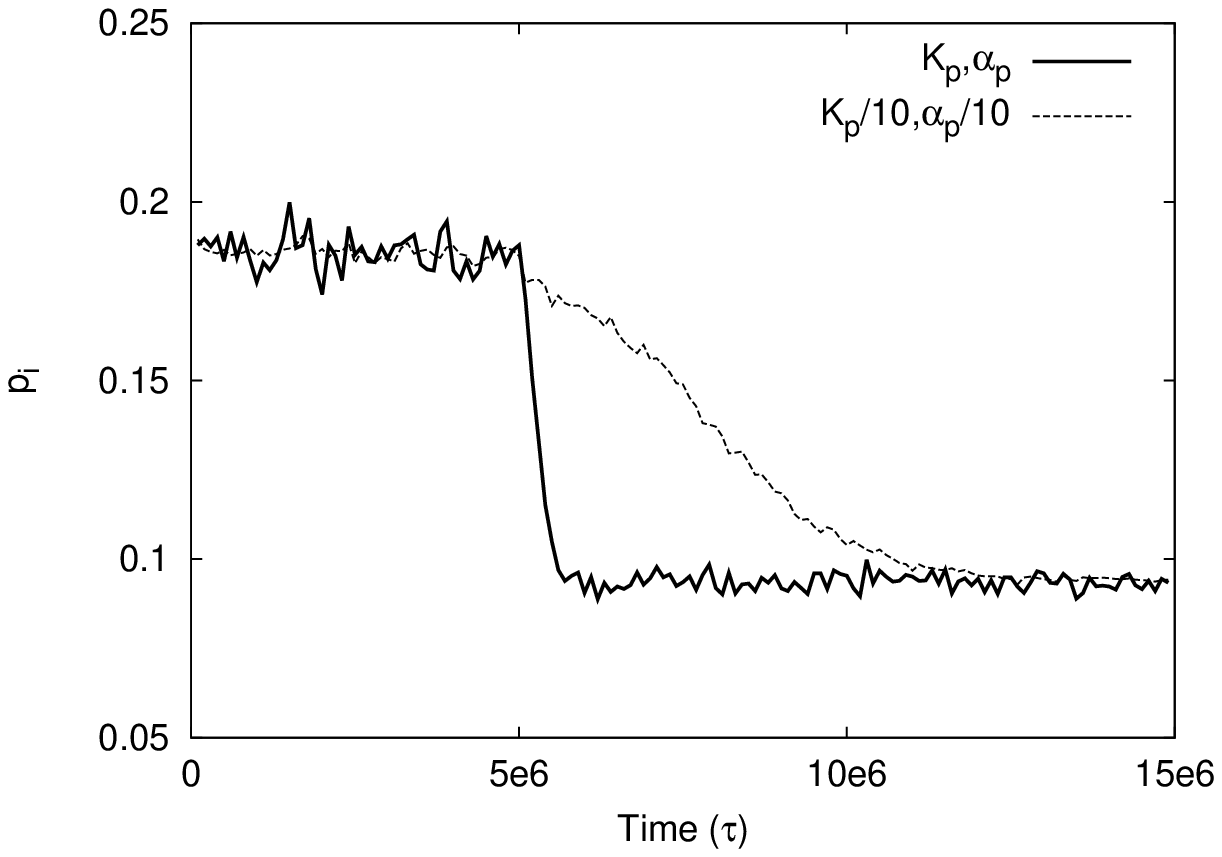}
	    \label{fig:reaction_pi}
        }%
        \subfloat[]{
                \centering
	    \includegraphics[width=0.33\linewidth]{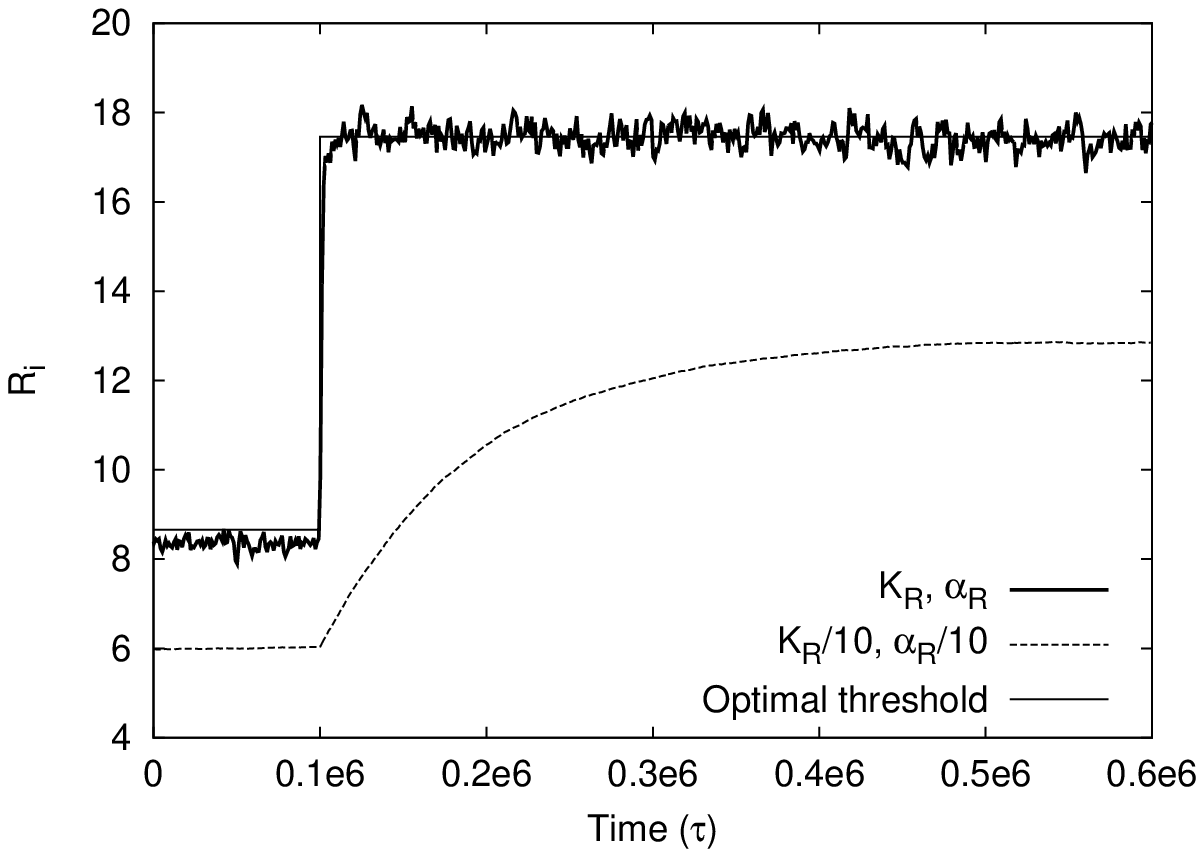}
	    \label{fig:reaction_Ri}
        }
        \subfloat[]{
                \centering
	    \includegraphics[width=0.33\linewidth]{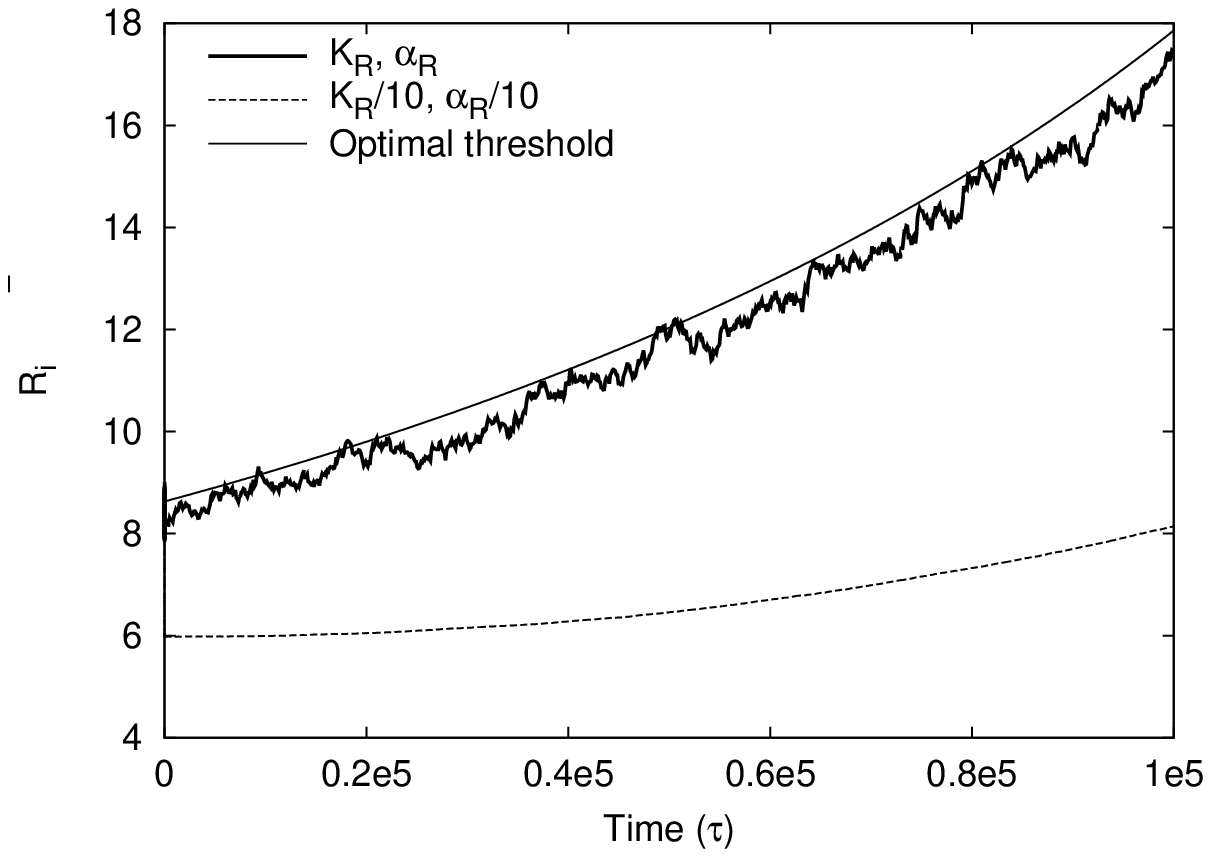}
	    \label{fig:speed}
        }        
          
  \caption{Speed of reaction when (a) a station joins the network, (b) there is a drastic change of $\rho$, and (c) a station is moving.}
  
  \label{fig:reaction}
\end{figure*}

To analyze the speed of reaction to changing radio conditions, we consider the following two scenarios: ($i$) a drastic change of the normalized SNR, which could be caused, e.g., by a \new{drastic change in weather conditions\cite{rain}}, and ($ii$) a soft change of the normalized SNR caused, e.g., by the movement of the station. Both scenarios trigger the adjustment of $\bar{R}_i$; in the following, we study the evolution of $\bar{R}_i$ in each case.
For the first scenario, we consider that in a wireless network with $N=2$ stations, both of them with a normalized SNR $\rho=1$, one of the stations changes its normalized SNR to $\rho=4$ after a time \new{$10^5 \tau$}. Fig.~\ref{fig:reaction_Ri} shows the evolution over time of the $\bar{R}_i$ of the station whose normalized SNR has changed, for the proposed setting of the $\{K_R,\alpha_R\}$ parameters as well as for a setting of these parameters 10 smaller. As a benchmark, the figure also shows the optimal setting of the threshold as given by the analytical results.
The results show that: ($i$) with our configuration, the system reacts quickly and closely follows the benchmark, while the reaction is much slower for a smaller setting;
and ($ii$) the steady state error with our setting is negligible, whereas with a smaller setting of the parameters it is much larger. The latter effect is caused by the fact that the steady error with a proportional controller increases as its proportional gain ($K_R$) is reduced. Therefore, by choosing a too small value for $K_R$, we do not only worsen the speed of reaction of the system but also its steady error.
For the second scenario, we consider a station moving towards the receiver at a constant speed: initially, the station is located at a distance $D$ (with an average normalized SNR of $\rho = 1$) and it moves to a distance $D/2$ of the sending station over a period of $10^5 \tau$.\footnote{Note that this corresponds to a high speed -- we cover half of the distance to the destination in $100ms$ if $\tau = 1\mu s$.} We consider a path loss exponent equal to 2. Fig.~\ref{fig:speed} shows the evolution of the $\bar{R}_i$ of the moving station over time. We observe that with our setting, the algorithm that adjusts $\bar{R}_i$ is able to cope with the movement of the station and $\bar{R}_i$ closely follows the optimal threshold. As in the previous case, with a smaller setting of the parameters the threshold used is far from the optimal because of the slow speed of reaction as well as the steady error.

\new{These results illustrate that the configuration  of the parameters $\{K_p,\alpha_p\}$ and $\{K_R,\alpha_R\}$ proposed in \S\ref{sec-analysis}} provides a good tradeoff between stability, speed of reaction and steady error: with a larger setting of the parameters the system suffers from instability, while with a smaller setting, it reacts too slowly and yields a large steady error.






%

\subsection{Moving Stations}\label{sec:perf-moving}

While the previous experiment involved only a single mobile station, in many cases some or all of the terminals 
may be moving. We next investigate a more complex 
scenario where stations move in an area of size $L$x$L$ following the random waypoint model, and send data to a station located at position ($L$,$L$). The transmission power is such that the normalized SNR for a station located at position (0,0) is $\rho = 1$.\footnote{Note that we do not let $\rho$ increase any further once a station is than a distance of $L/100$ to the receiver.}
We further consider a path loss exponent equal to~2. 
We compare ADOS against the following approaches: ($i$) a benchmark that uses, for the current normalized SNR, the optimal transmission rate threshold obtained from the analytical results (`\emph{optimal}'), ($ii$) the `\emph{non-opportunistic}' approach, \new{($iii$) CSMA/CA}, ($iv$) TDOS, ($v$) NDOS, and ($vi$) the approach we proposed in \cite{ourinfocom} (`\emph{static} ADOS').\footnote{Note that the approach proposed in this paper differs from the previous conference version \cite{ourinfocom} in that it adapts to changing radio conditions; therefore, when radio conditions are static (as in experiments \ref{sec-perf-first} to \ref{sec-perf-last}) both behave in the same way.} For the TDOS, NDOS and `\emph{static} ADOS' approaches, since they assume static radio conditions and hence rely on long term measurements to set the transmission rate threshold, we measure the average SNR over periods of $10^{8}\tau$ and use the measurement obtained in a period to compute the $\bar{R}_i$ of the next period. \new{This corresponds to a large window of time, to model the static nature of those schemes.}

\begin{figure*}[t!]        

\centering
             \subfloat[]{
	    \includegraphics[width=0.33\linewidth ]{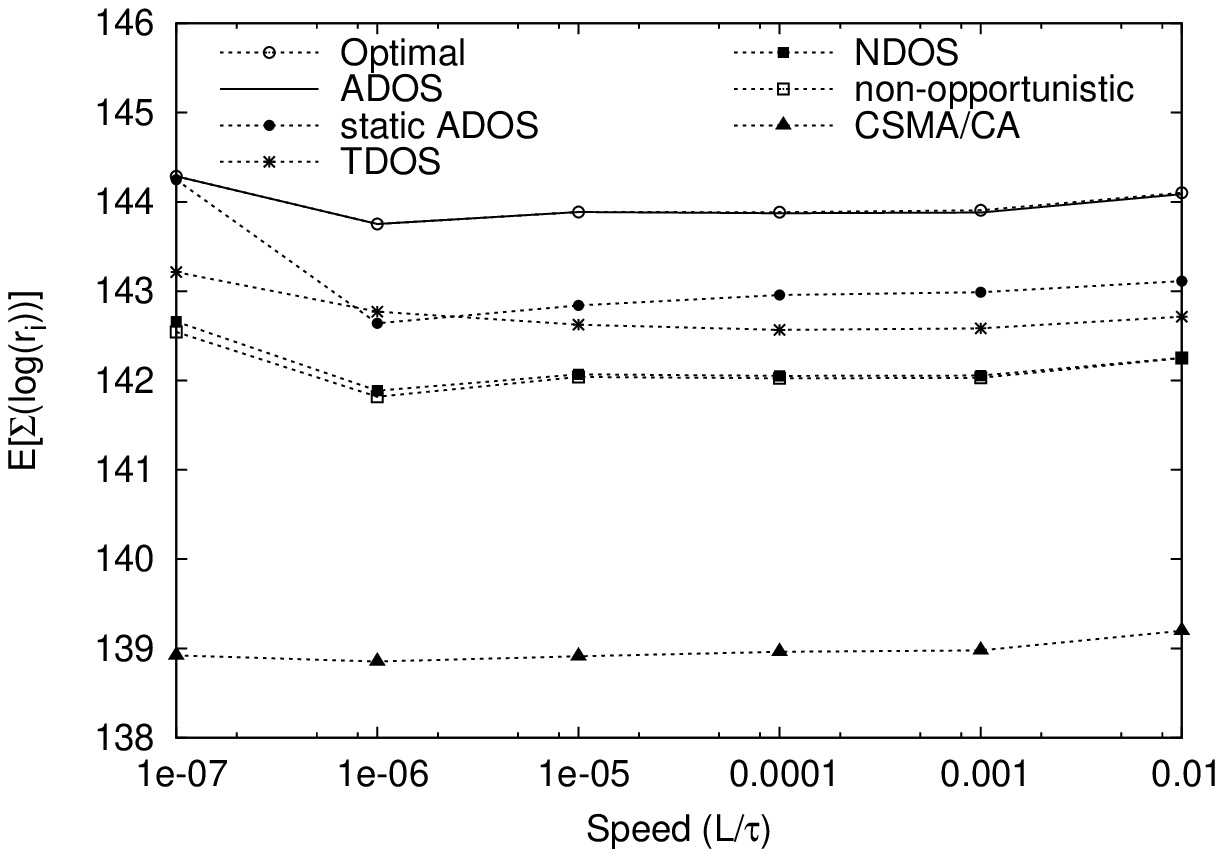}
	    \label{fig:moving}
        }%
        \subfloat[]{
            \centering
	    \includegraphics[width=0.66\linewidth]{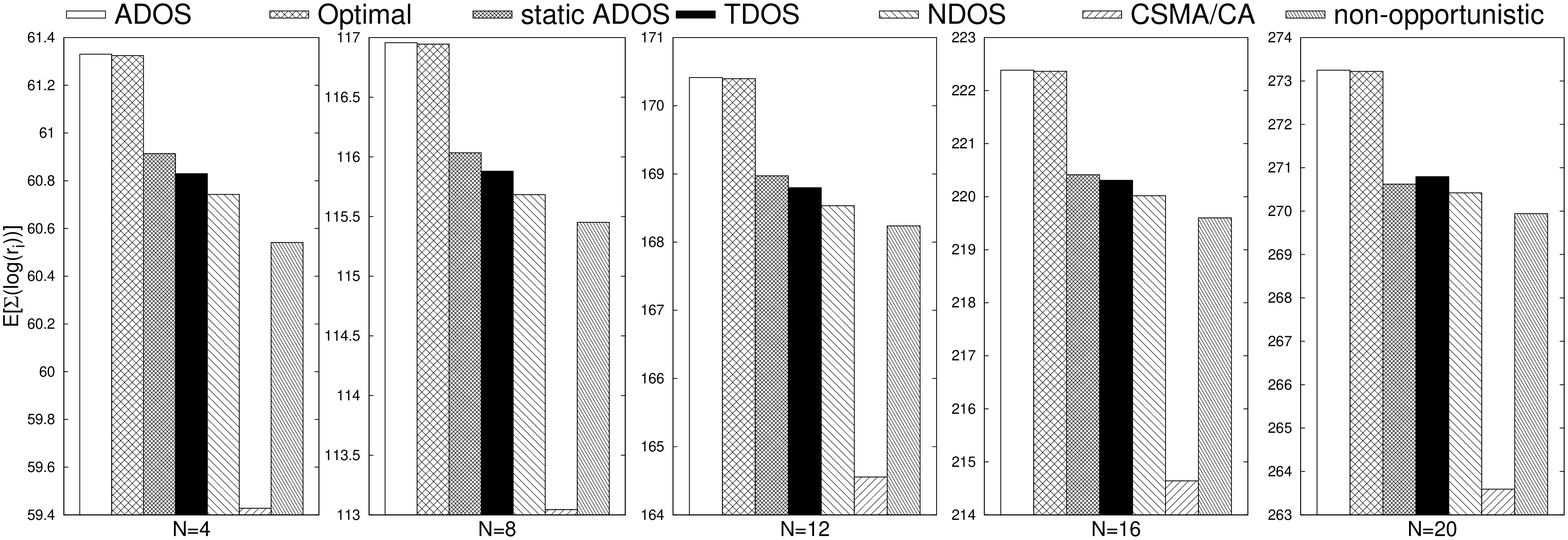}

	    \label{fig:mobility_nodes}
        }
  
  \caption{\new{Moving stations. $\sum_i{\log(r_i)}$ (a)  vs. speed for $N=10$; and (b) vs. number of stations for a speed of $10^{-5}$ L/$\tau$.  }}
  
  \label{fig:other_channel}
\end{figure*}

\new{We evaluate their performance as a function of the speed (in units of $L/\tau$) in Fig.~\ref{fig:moving} and as a function of the number of stations in Fig.~\ref{fig:mobility_nodes}}, in terms of the $\sum_i{\log(r_i)}$ averaged over intervals of $10^4\tau$. By averaging the $\sum_i{\log(r_i)}$ over different time intervals, we not only capture the long-term fairness (i.e., fairness in total throughput) but also the short-term fairness (i.e., fairness in the throughput obtained over a given time interval).\footnote{Note that in the previous experiments where radio conditions were static, short-term fairness was not an issue.} 
We observe from the results that the performance of ADOS closely follows the `\emph{optimal}' benchmark 
and outperforms all other approaches. 
As in previous experiments, the `\emph{non-opportunistic}', \new{CSMA/CA}, TDOS and NDOS approaches perform substantially worse that ADOS. 
The `\emph{static} ADOS' approach also performs substantially worse, as it does not adjust to current radio conditions. 
While it does perform well for very low speeds for which the measurement period is sufficient, performance degrades sharply when the speed increases and stations far from the destination with an outdated threshold risk starvation. Performance improves slightly for even higher speeds, as the probability that a station stays far from the destination during the entire averaging period decreases, i.e., the threshold is outdated but due to the high speed the station is often near the destination. \new{In contrast, the relative gains of ADOS are practically independent of the number of stations.}

\section{\new{Implementation issues and challenges}} \label{sec-practical}
\new{
One of the main advantages of ADOS is that it only relies on local information already available in commodity hardware. This simplifies its implementation as stations are neither required to exchange information nor to carry out complex operations in contrast to other approaches, like that of \cite{itit} that requires both a protocol to exchange network information and to solve definite integrals. 
ADOS employs a contention-based MAC protocol similar to the CSMA/CA protocol used in  IEEE 802.11, which also divides time into slots and implements a random channel access. When contending, IEEE 802.11 uses a binary exponential backoff algorithm~\cite{bianchi} based on a ``contention window'' number ($cw$)  that is initialized to a minimum $cw_{min}$ every successful access and it is doubled every failed attempt, until a $cw_{max}$ is reached. 
We are implementing ADOS through simple modifications to the kernel module {\ttfamily mac80211},\footnote{\new{{\ttfamily\url{http://wireless.kernel.org/en/developers/Documentation/mac80211/}}}} which is common to all 802.11 platforms in the Linux wireless stack, and minor modifications to the {\ttfamily  openFWWF} firmware\footnote{\new{{\ttfamily\url{http://www.ing.unibs.it/~openfwwf/}}}}  of IEEE 802.11 Broadcom cards for time-sensitive operations related to the channel probing mechanism. We note that a proper experimental evaluation  requires controlled fading environments e.g., using a channel emulator, and thus we leave it as future work. These modifications are summarized as follows.

\subsubsection{Channel probing}

A simple solution to deploy ADOS' channel probing over an 802.11 stack is to use the standard RTS/CTS mechanism of IEEE 802.11, available in practically all platforms. With this mechanism, a station willing to transmit sends first an RTS message to the intended receiver which in turn replies with a CTS response; only upon reception of this CTS message, the transmitter can send its data. In this way, all the neighbouring stations are aware of the ongoing communication. From the RTS/CTS it is possible to extract the RSSI (Received Signal Strength Indicator)\footnote{\new{We are also exploring other alternatives, such as exploiting the Channel State Information (CSI), a standard feedback structure that provides a much richer source of link quality information than RSSI~\cite{halperin2011predictable}.}}, a measure of the link quality that we can exploit  to decide whether to skip a transmission opportunity or not, and the rate to use in that case (as we describe next).

\subsubsection{Rate adaptation and threshold}

 Upon the reception of an RTS, the receiver uses this message's channel information to compute the optimal modulation and coding scheme (MCS) for the transmitter; if the bit rate provided by such MCS  falls below the rate threshold $\bar{R}_i$, it does reply with a CTS setting  the \emph{duration} field\footnote{\new{The \emph{duration} field, included in the MAC header of IEEE 802.11 frames, is used by the virtual carrier-sensing mechanism NAV (Network Allocation Vector) to advertise the amount of time the medium will be busy so others do not contend.}}  to 0 to announce the transmitter (and the other overhearing stations) that there shall be no communication and all nodes can re-contend. If the estimated bit rate is larger than the threshold, the station embeds such MCS within the CTS message\footnote{\new{The authors of \cite{halperin2011predictable} follow a similar idea, embedding such info into ACKS.}} and sends it back to the transmitter with the \emph{duration} field set to the frame length (so overhearing stations do not contend during this time). To compute the optimal MCS, each station keeps track of the link quality when receiving data from every other station, and implements a rate adaptation algorithm to compute the best MCS (there is a plethora of algorithms available that are easily deployable over off-the-shelf devices, e.g., \cite{camp2010modulation, zhang2008practical, halperin2011predictable}).


\subsubsection{Channel sensing, contention parameters and frame construction}
A preliminary modification to the RTS/CTS mechanism is to set the \emph{duration} field of the RTS message to only the duration of the RTS/CTS exchange. This provides enough protection to our probe scheme (i.e., no contention occurs while a station is probing the channel) and permits re-contention if the threshold is not satisfied. In turn,
ADOS'  access probability $p_i$ can be set, regardless of the number of stations, by computing $cw_i=\frac{2}{p_i} -1$ and setting  $cw_{i,min}=cw_{i,max}=cw_i$~\cite{bianchi2000performance}.
Finally, the fixed frame duration of ADOS can be  set in {\ttfamily mac80211} by implementing a \emph{leaky bucket} controller that limits the frame size to $L_i(t) \approx \left(\mathcal{T}-SIFS-T_{ACK}\right ) R_i(t)$, where $SIFS$ is the interval of time between data transmission and acknowledgment  reception (ACK), $T_{ACK}$ is the duration of an ACK, and $R_i(t)$ is the bit rate  selected for this frame.

}

\section{Conclusions}\label{sec-conclusions}

Distributed Opportunistic Scheduling (DOS) techniques provide throughput gains in wireless networks without requiring a centralized scheduler. One of the challenges of these techniques is the design of an adaptive algorithm that adjusts the DOS parameters to their optimal value. In this paper we propose a novel algorithm, named ADOS, with the following advantages: ($i$) it jointly optimizes both the access probabilities and the transmission thresholds; ($ii$) it provides a good tradeoff between total throughput and fairness; and ($iii$) it guarantees convergence and stability. A major finding when computing the configuration of the optimal threshold is that it is independent of the access probabilities, which \new{allows us to design two independent mechanisms to compute thresholds and access probabilities, respectively.}
The performance of ADOS has been extensively evaluated via simulations. Results confirm that ADOS provides significantly better performance than previous proposals; in particular, key results are that ADOS outperforms other approaches substantially with non-saturated stations as well as with changing radio conditions.


\begin{thebibliography}{10}
\providecommand{\url}[1]{#1}
\csname url@samestyle\endcsname
\providecommand{\newblock}{\relax}
\providecommand{\bibinfo}[2]{#2}
\providecommand{\BIBentrySTDinterwordspacing}{\spaceskip=0pt\relax}
\providecommand{\BIBentryALTinterwordstretchfactor}{4}
\providecommand{\BIBentryALTinterwordspacing}{\spaceskip=\fontdimen2\font plus
\BIBentryALTinterwordstretchfactor\fontdimen3\font minus
  \fontdimen4\font\relax}
\providecommand{\BIBforeignlanguage}[2]{{%
\expandafter\ifx\csname l@#1\endcsname\relax
\typeout{** WARNING: IEEEtran.bst: No hyphenation pattern has been}%
\typeout{** loaded for the language `#1'. Using the pattern for}%
\typeout{** the default language instead.}%
\else
\language=\csname l@#1\endcsname
\fi
#2}}
\providecommand{\BIBdecl}{\relax}
\BIBdecl

\bibitem{ourinfocom}
A.~Garcia-Saavedra, A.~Banchs, P.~Serrano, and J.~Widmer, ``Distributed
  opportunistic scheduling: A control theoretic approach,'' in
  \emph{Proceedings of IEEE INFOCOM}, Orlando, FL, March 2012.

\bibitem{cao2006}
M.~Cao, V.~Raghunathan, and P.~Kumar, ``Cross-layer exploitation of mac layer
  diversity in wireless networks,'' in \emph{Proceedings of IEEE ICNP}, Santa
  Barbara, CA, November 2006.

\bibitem{asadi2013survey}
A.~Asadi and V.~Mancuso, ``A survey on opportunistic scheduling in wireless
  communications,'' \emph{IEEE Communications Surveys \& Tutorials}, vol.~15,
  no.~4, pp. 1671--1688, 2013.

\bibitem{ghosh2009priority}
C.~Ghosh, S.~Chen, D.~P. Agrawal, and A.~M. Wyglinski, ``Priority-based
  spectrum allocation for cognitive radio networks employing nc-ofdm
  transmission,'' in \emph{Military Communications Conference, 2009. MILCOM
  2009. IEEE}.\hskip 1em plus 0.5em minus 0.4em\relax IEEE, 2009, pp. 1--5.

\bibitem{itit}
D.~Zheng, W.~Ge, and J.~Zhang, ``Distributed opportunistic scheduling for ad
  hoc networks with random access: an optimal stopping approach,'' \emph{IEEE
  Transactions on Information Theory}, vol.~55, no.~1, January 2009.

\bibitem{ton}
P.~Thejaswi \emph{et~al.}, ``Distributed opportunistic scheduling with
  two-level probing,'' \emph{IEEE/ACM Transactions on Networking}, vol.~18,
  no.~5, October 2010.

\bibitem{twc}
D.~Zheng,  \emph{et~al.}, ``Distributed opportunistic scheduling for ad hoc
  communications with imperfect channel information,'' \emph{IEEE Transactions
  on Wireless Communications}, vol.~7, no.~12, December 2008.

\bibitem{infocom}
S.~Tan, D.~Z.~J. Zhang, and J.~R. Zeidler, ``Distributed opportunistic
  scheduling for ad-hoc communications under delay constraints,'' in
  \emph{Proceedings of IEEE INFOCOM}, San Diego, CA, March 2010.

\bibitem{dosjsac}
H.~Chen and J.~Baras, ``{Distributed Opportunistic Scheduling for Wireless
  Ad-Hoc Networks with Block-Fading Model},'' \emph{IEEE Journal on Selected
  Areas in Communications}, November 2013.

\bibitem{capacity-2014}
J.~Kampeas, A.~Cohen, and O.~Gurewitz, ``Capacity of distributed opportunistic
  scheduling in non-homogeneous networks,'' \emph{Information Theory, IEEE
  Transactions on}, vol.~PP, no.~99, pp. 1--1, 2014.

\bibitem{dos-hybrid}
W.~Mao, S.~Wu, and X.~Wang, ``Qos-oriented distributed opportunistic scheduling
  for wireless networks with hybrid links,'' in \emph{Globecom Workshops (GC
  Wkshps), 2013 IEEE}, Dec 2013, pp. 4524--4529.

\bibitem{Kelly}
F.~Kelly, ``{Charging and rate control for elastic traffic},'' \emph{European
  Transactions on Telecommunications}, vol.~8, pp. 33--37, 1997.

\bibitem{doc}
A.~Banchs, A.~Garcia-Saavedra, P.~Serrano, and J.~Widmer, ``A game-theoretic
  approach to distributed opportunistic scheduling,'' \emph{Networking,
  IEEE/ACM Transactions on}, vol.~21, no.~5, pp. 1553--1566, Oct 2013.

\bibitem{mobicom02}
B.~Sadhegi, V.~Kanodia, A.~Sabharwal, and E.~Knightly, ``Opportunistic media
  access for multirate ad hoc networks,'' in \emph{Proceedings of ACM MOBICOM},
  Atlanta, GA, September 2002.

\bibitem{infocom05}
P.~Gupta, Y.~Sankarasubramaniam, and A.~Stolyar, ``Random-access scheduling
  with service differentiation in wireless networks,'' in \emph{Proceedings of
  IEEE INFOCOM}, Miami, FL, March 2005.

\bibitem{birgitta}
B.~Kristiansson and B.~Lennartson, ``{Robust Tuning of PI and PID
  Controllers},'' \emph{IEEE Control Systems Magazine}, vol.~26, no.~1, pp.
  55--69, February 2006.

\bibitem{1204884}
A.~K. Palit and D.~Popovic, \emph{Computational Intelligence in Time Series
  Forecasting: Theory and Engineering Applications}.\hskip 1em plus 0.5em minus
  0.4em\relax Springer-Verlag New York, Inc., 2005.

\bibitem{franklin}
G.~F. Franklin, J.~D. Powell, and M.~L. Workman, \emph{{Digital Control of
  Dynamic Systems}}, 2nd~ed.\hskip 1em plus 0.5em minus 0.4em\relax
  Addison-Wesley, 1990.

\bibitem{boggia}
G.~Boggia, P.~Camarda, L.~A. Grieco, and S.~Mascolo, ``Feedback-based control
  for providing real-time services with the 802.11e mac,'' \emph{IEEE/ACM
  Transactions on Networking}, vol.~15, no.~2, April 2007.

\bibitem{tmc2}
A.~Banchs, P.~Serrano, and L.~Vollero, ``{Providing Service Guarantees in
  802.11e EDCA WLANs with Legacy Stations},'' \emph{IEEE Transactions on Mobile
  Computing}, vol.~9, no.~8, pp. 1057--1071, August 2010.

\bibitem{hollow}
C.~V. Hollot, V.~Misra, D.~Towsley, and W.~B. Gong, ``{A Control Theoretic
  Analysis of RED},'' in \emph{Proceedings of IEEE INFOCOM}, Anchorage, Alaska,
  April 2001.

\bibitem{weak-strong-stability}
\BIBentryALTinterwordspacing
E.~Leonardi, ``\BIBforeignlanguage{English}{Throughput optimal scheduling
  policies in networks of constrained queues},''
  \emph{\BIBforeignlanguage{English}{Queueing Systems}}, vol.~78, no.~3, pp.
  197--223, 2014. [Online]. Available:
  \url{http://dx.doi.org/10.1007/s11134-014-9407-9}
\BIBentrySTDinterwordspacing

\bibitem{astrom}
K.~Astr\"om and B.~Wittenmark, \emph{{Computer-controlled systems, theory and
  design}}, 2nd~ed.\hskip 1em plus 0.5em minus 0.4em\relax Prentice Hall
  International Editions, 1990.

\bibitem{jain}
R.~Jain, D.~M. Chiu, and W.~Hawe, ``A quantitative measure of fairness and
  discrimination for resource allocation in shared computer systems,'' DEC,
  Tech. Rep. TR-301, 1984.

\bibitem{jakes}
W.~C. Jakes, \emph{Microwave Mobile Communications}.\hskip 1em plus 0.5em minus
  0.4em\relax New York: John Wiley \& Sons Inc., 1975.

\bibitem{rain}
R.~Crane, ``Prediction of attenuation by rain,'' \emph{Communications, IEEE
  Transactions on}, vol.~28, no.~9, pp. 1717--1733, Sep 1980.

\bibitem{bianchi}
G.~Bianchi, ``{Performance Analysis of the IEEE 802.11 Distributed Coordination
  Function},'' \emph{IEEE Journal on Selected Areas in Communications},
  vol.~18, no.~3, pp. 535--547, March 2000.

\bibitem{halperin2011predictable}
D.~Halperin, W.~Hu, A.~Sheth, and D.~Wetherall, ``Predictable 802.11 packet
  delivery from wireless channel measurements,'' \emph{ACM SIGCOMM Computer
  Communication Review}, vol.~41, no.~4, pp. 159--170, 2011.

\bibitem{camp2010modulation}
J.~Camp and E.~Knightly, ``Modulation rate adaptation in urban and vehicular
  environments: cross-layer implementation and experimental evaluation,''
  \emph{Networking, IEEE/ACM Transactions on}, vol.~18, no.~6, pp. 1949--1962,
  2010.

\bibitem{zhang2008practical}
J.~Zhang, K.~Tan, J.~Zhao, H.~Wu, and Y.~Zhang, ``A practical snr-guided rate
  adaptation,'' in \emph{INFOCOM 2008. The 27th Conference on Computer
  Communications. IEEE}.\hskip 1em plus 0.5em minus 0.4em\relax IEEE, 2008.

\bibitem{bianchi2000performance}
G.~Bianchi, ``Performance analysis of the ieee 802.11 distributed coordination
  function,'' \emph{Selected Areas in Communications, IEEE Journal on},
  vol.~18, no.~3, pp. 535--547, 2000.

\end{thebibliography}
\end{document}